\documentclass[twocolumn,english,aps,pre,showpacs,nofootinbib]{revtex4}
\usepackage[T1]{fontenc}
\usepackage[latin9]{inputenc}
\usepackage{amsmath}
\usepackage{graphicx}
\usepackage{amssymb}
\usepackage{esint}
\usepackage[usenames]{color}
\usepackage{hyperref}

\makeatletter
\usepackage[normalem]{ulem}\usepackage{float}

\usepackage{ulem}\usepackage{amsfonts}

\makeatother

\usepackage{babel}

\makeatother

\usepackage{babel}

\begin{document}

\title{Nonlinear effective medium theory of disordered spring networks}

\author{M. Sheinman}
\affiliation{Department of Physics and Astronomy, Vrije Universiteit, Amsterdam, The Netherlands}
\affiliation{Kavli Institute for Theoretical Physics, University of California, Santa Barbara, California 93106, USA}

\author{C. P. Broedersz}
\affiliation{Department of Physics and Astronomy, Vrije Universiteit, Amsterdam, The Netherlands}
\affiliation{Kavli Institute for Theoretical Physics, University of California, Santa Barbara, California 93106, USA}

\author{F. C. MacKintosh}
\affiliation{Department of Physics and Astronomy, Vrije Universiteit, Amsterdam, The Netherlands}
\affiliation{Kavli Institute for Theoretical Physics, University of California, Santa Barbara, California 93106, USA}

\date{\today}
\begin{abstract}
\noindent Disordered soft materials, such as fibrous networks in biological contexts exhibit a nonlinear elastic response. We study such nonlinear behavior with a minimal model for networks on lattice geometries with simple Hookian elements with disordered spring constant. By developing a  mean-field approach to calculate the differential elastic bulk modulus for the macroscopic network response of such networks under large isotropic deformations, we provide insight into the origins of the strain stiffening and softening behavior of these systems. We find that the nonlinear mechanics depends only weakly on the lattice geometry and is governed by the average network connectivity. In particular, the nonlinear response is controlled by the isostatic connectivity, which depends strongly on the applied strain. Our predictions for the strain dependence of the isostatic point as well as the strain-dependent differential bulk modulus agree well with numerical results in both two and three dimensions. In addition, by using a mapping between the disordered network and a regular network with random forces, we calculate the non-affine fluctuations of the deformation field and compare it to the numerical results. Finally, we discuss the limitations and implications of the developed theory.
\end{abstract}
\maketitle

%\tableofcontents{}
\section{Introduction}

\noindent Rich elastic behavior is a common feature of many soft materials such as foams,
granular packings and soft glasses \cite{liu1998jamming,cates1998jamming}, as well as networks of protein fibers that form major
structural components of cells and tissue \cite{bausch2006bottom,kasza2007cell,fletcher2010cell}.
One characteristic these varied systems share is their particular sensitivity to external stress;
in densely jammed systems, for instance, the external pressure can cause the system
to transition between rigid and floppy states \cite{cates1998jamming,liu1998jamming,o2003jamming,wyart2008,van2010jamming,liu2010jamming},
while reconstituted biological filamentous networks
exhibit dramatic strain stiffening under shear~\cite{janmey1991viscoelastic,gardel2004elastic,storm2005nonlinear,lieleg2007mechanics}.
This remarkable nonlinear elastic behavior of fiber networks has attracted much attention in the last decade;
in addition to the physiological relevance of this nonlinear elastic response for cells and many biological tissues
\cite{shadwick1999mechanical,winer2009non}, these systems are also interesting from a fundamental perspective,
owing to their unusual nonlinear materials properties~\cite{gardel2004elastic,storm2005nonlinear,onck2005alternative,gardel2006prestressed,
wagner2006cytoskeletal,huisman2007three,heussinger2007nonaffine,lieleg2007mechanics,
chaudhuri2007reversible,broedersz2008nonlinear,kasza2009nonlinear,conti2009cross}, including negative normal stresses \cite{janmey2006negative}. Understanding how their intrinsic disordered nature affects the elastic deformations is required for a complete theoretical description of their nonlinear mechanical behavior. Although structural disorder and inhomogeneous deformations clearly play a central role in jamming systems \cite{bernal1960packing,cates1998jamming,wyart2008}, their precise role in the nonlinear behavior of fibrous networks remains unclear \cite{onck2005alternative,huisman2007three,kabla2007nonlinear,heussinger2007nonaffine,wyart2008,conti2009cross,
chen2010rheology}.

Prior work on the nonlinear elasticity of random spring networks has focussed on triangular lattices with internal stresses~\cite{tang1988percolation}. In the limit of small disorder, a perturbation theory was applied to describe the nonlinear elastic response of such systems with \emph{small} dilution. It was shown numerically that the transition value of the mean coordination number, at which the network acquires rigidity, shifts with applied strain. Interestingly, the perturbation theory also appeared to capture the behavior observed numerically even for highly diluted networks, since the bulk modulus was found to increase linearly with the mean coordination number beyond the rigidity percolation point. Recently, similar nonlinear behavior was analysed for random spring networks in jammed configurations. Consistent with prior work on triangular lattices~\cite{tang1988percolation}, it was shown that this nonlinear response is controlled by the central force isostatic point~\cite{wyart2008}; this isostatic point characterizes the average connectivity, $z$, at which the number of central-force constraints balances the number of degrees of freedom in the system and is given by $z_0=2d$ in $d$ dimensions \cite{Maxwell1864}. For jammed systems, it was shown that the nonlinear response close to this isostatic point is well described by a mean field scaling approach~\cite{wyart2008}. By contrast, the systems we consider here are not in jammed configurations, but instead fall in the class of lattice-based rigidity percolation problems \cite{thorpe1983continuous,feng1984,jacobs1996}. For instance, the linear elastic response of fiber networks is also governed by the central-force isostatic point for a broad range of network connectivities---even with fibers with non-central fiber bending interactions---but with non-mean-field behavior~\cite{Broedersz2011}. Motivated by these results, we investigate the nonlinear behavior of fiber networks in the limit of vanishing bending rigidity under large deformations.

\begin{figure}
\includegraphics[width=\columnwidth]{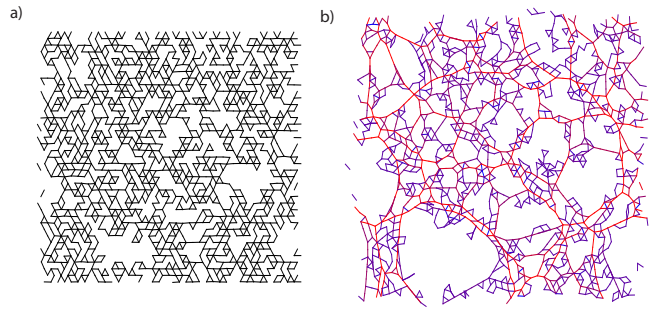}

\caption{(Color online) A small section of the relaxed (left) and expanded (right) diluted triangular lattice. The average coordination number in this example is $z=3$. On the right plot the blue/red colors mixture represents the high/low elastic energy stored in a bond. See \url{http://www.youtube.com/watch?v=ANSQePygfYU} for a stiffening movie.}
\label{Expansion}
\end{figure}

\begin{figure}[tp]
\includegraphics[width=\columnwidth]{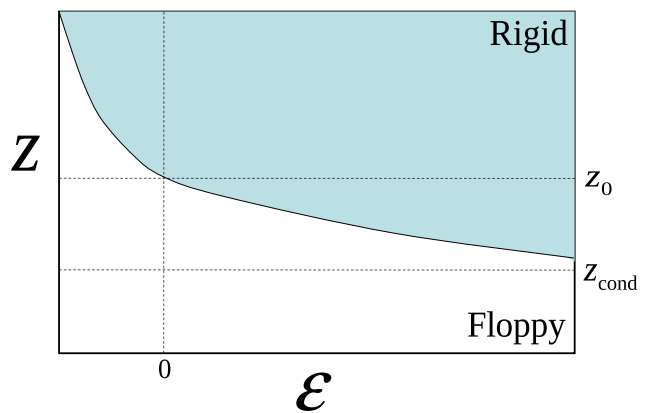}
\caption{(Color online) The schematic phase diagram for the rigidity of random spring networks under an isotropic strain $\epsilon$. The central-force isostatic point, $z_{0}$, the conductivity threshold, $z_{\rm cond}$ and the lower rigidity threshold in negative strain are indicated in the diagram.
}
\label{phasescheme}
\end{figure}

% In the Summary Section we discuss in more detail the relevance of the proposed model and its mean-field
%solution for the real biopolymer gels with fiber bending and other related systems such as network glasses and jammed systems.
%
Here we investigate the nonlinear elastic response of random spring networks under isotropic expansion or compression over a broad range of network connectivities---both above and below the small strain isostatic point $z_0$, as illustrated in Fig. \ref{Expansion}. From simulations we find that disordered \emph{subisostatic} spring networks exhibit significant strain-stiffening. We gain insight into the origins of this behavior by developing an effective medium theory (EM theory) for the nonlinear responses of random spring networks on lattice geometries. The nonlinear behavior of these systems depends only weakly on network geometry and appears to be controlled largely by the mean network connectivity, $z$, and the applied strain, $\epsilon$.  Within the framework of this central-force network model, the network's stiffness exhibits a transition on the two-dimensional phase diagram in $\epsilon$ and $z$, which characterizes the strain dependence of the isostatic point, a shown in Fig.~\ref{phasescheme}. The transition curve, $z_c\left(\epsilon\right)$---representing the transition between a rigid and a floppy state---is derived using the EM theory and found to agree well with the numerical results. Interestingly, the mean-field solution predicts that a \emph{superisostatic} central-force network looses rigidity and collapses beyond a threshold in compressional strain, as was observed for perfect two-dimensional triangular network in Monte-Carlo simulations at low temperatures \cite{discher1997phase,wintz1997mesh}. The theoretical predictions are verified using numerical calculation.

The mentioned above results for random spring networks may lend insight in the mechanics of biopolymer networks. The nonlinear elasticity of reconstituted networks of intracellular biopolymers such as filamentous actin (F-actin) and intermediate filaments has in many cases
been accounted for by the affine entropic model \cite{mackintosh1995elasticity,gardel2004elastic,storm2005nonlinear,lin2010origins}. In this model, network disorder is ignored by assuming a uniform (affine) deformation and, consequently, the nonlinear network response is directly determined by the nonlinear entropic force-extension behavior of the individual filaments. By contrast, there is increasing evidence that the strain-stiffening behavior of networks consisting of stiff thick fibers, such as collagen and bundled actin networks is governed by collective non-affine fiber bending deformations~\cite{lieleg2007mechanics,stein2008micromechanics,chen2010rheology,Lindstrom2010biopolymer}. Despite extensive analytical and numerical investigations \cite{wilhelm2003elasticity,head2003distinct,head2003deformation,onck2005alternative,huisman2007three, kabla2007nonlinear,heussinger2007nonaffine,heussinger2007statistical,van2008models,wyart2008,conti2009cross,broedersz2011molecular}, the principles of such network deformations and the resulting nonlinear network response are still unclear. To investigate the implications of our results on random spring networks for biological filamentous networks, we include a finite bending rigidity for the fibers in our network model. For, sufficiently small bending rigidities, these networks exhibit nonlinear elastic behavior. However, in this case the nonlinear behavior is not due to a transition between a floppy and a rigid state, but between a soft bending-dominated elastic behavior and a stiffer stretching-dominated behavior~\cite{onck2005alternative}. Both with and without bending rigidity, the nonlinear response is still governed by the central force isostatic point.

The outline of the paper is as follows. In Sec. \ref{The model} we define the model and summarize the applied approach and the main results of the paper. In Sec. \ref{Effective medium theory} we present the mean-field approach in detail, derive the differential bulk modulus of a system, analyze the non-affine fluctuations and, by using a self-consistency check, we identify the range of applicability of the performed approximations. In Sec. \ref{Diluted regular networks} we demonstrate the presented general method using a particular example of the diluted regular networks and compare the analytical predictions to the numerical results. We discuss the results and their implications and summarize in Sec. \ref{Summary and discussion}.
\section{The model}\label{The model}
In this paper we analyze the nonlinear elastic behavior of a random central-force network on lattice-geometries with varying connectivity. We start out with a model in which the bending energy of the fibers or bonds is ignored. This model will allow us to study the effects of finite strain on the central-force isostatic point, $z_{c}$, and the stretching
energies of the bonds. To further simplify our model we only consider isotropic expansional and compressional strains (see Fig. \ref{Expansion}). The calculation of the elastic properties under nonlinear shear is complicated by the broken symmetry and is described elsewhere \cite{sheinman2012NonlinearShear}.
The network is constructed on an ordered lattice geometry in $d\geq1$ dimensions. We capture
the effects of disorder by a distribution
of the spring constants associated to the bonds in the network. In
this model the rest lengths of all springs are chosen to be identical
and equal to the lattice spacing, $\ell_{0}$.

Measuring all lengths in units of $\ell_{0}$, the Hamiltonian of
the system is given by
\begin{equation}
H=\frac{1}{2}\underset{\left\langle ij\right\rangle }{\sum}\mu_{ij}\left(|\mathbf{u}_{i}-\mathbf{u}_{j}|-1\right)^{2},
\label{Model}
\end{equation}
 where $\mathbf{u}_{i}$ is the position of vertex $i$, $\left\langle ij\right\rangle $
denotes the summation over neighbouring lattice vertices and $\mu_{ij}$
is the stretching modulus for the bond between vertices $i$ and $j$.
The stretching moduli $\mu_{ij}$ are drawn independently from a known probability
density, $P\left(\mu_{ij}\right)$.

\subsection{Quantities of interest}
Here, we investigate the
elastic response of the network to an applied global expansion/compression
strain,
\begin{equation}
\epsilon=\frac{L'-L}{L},
\end{equation}
where $L'$ and $L$ are the linear size of the strained and unstrained networks, respectively.
To quantify the nonlinear elastic response to the global bulk strain, we define the \textit{nonlinear differential bulk modulus} as
\begin{equation}
B\left(\epsilon\right)\equiv\frac{n}{d^2}\frac{\partial^{2}E\left(\epsilon\right)}{\partial\epsilon^{2}},
\label{K_def}
\end{equation}
where $E\left(\epsilon\right)$ is the average elastic energy per bond, $n$ is the density of bonds in the unstrained and undiluted lattice and $d$ is dimension of the system.
For example, for the FCC lattice $n=2\sqrt{6}\ell_0^{-3}$, while for the triangular lattice $n=2\sqrt{3}\ell_0^{-2}$.

This definition of the nonlinear bulk modulus has the following advantages:
\paragraph{}
Other quantities, related to the nonlinear response of the system to a global strain may be deduced from $B\left(\epsilon\right)$. For instance, the pressure,
\begin{equation}
\Pi=-\frac{\partial U}{\partial V},
\end{equation}
where $U=NE$ is the total elastic energy, $N$ is the total number of springs in the undiluted network, $V=V_0\left(1+\epsilon\right)^d$ is the system's volume and $V_0$ is the total volume of the unstrained network. This pressure can be obtained directly from the nonlinear differential bulk modulus using
\begin{equation}
\Pi=-\frac{d}{\left(1+\epsilon\right)^{d-1}}\int_0^\epsilon B\left(\epsilon\right)d\epsilon.
\end{equation}
\paragraph{}
In the linear regime, $\epsilon\rightarrow0$, the \textit{nonlinear} bulk modulus converges to
\begin{equation}
B\left(\epsilon\rightarrow 0\right)=%-\mathcal{B}\equiv
V \frac{\partial^2 U}{\partial V^2}.
\end{equation}
\paragraph{}
If the material is composed of Hookian bonds and its deformation is affine, $B\left(\epsilon\right)$ is constant and equal to $n/d^2$ times the average spring constant of the network. Thus, by plotting $\frac{d^2}{n}B\left(\epsilon\right)$ one can easily compare the actual elastic properties to the affine predictions.

\subsection{Numerical results}\label{Numerical results}
To study the nonlinear elastic response of random spring networks, we choose a specific realization of the spring constant probability density for networks on lattice geometries (see Sec. \ref{Diluted regular networks} for a detailed discussion). In particular, we investigate bond diluted network of springs with a modulus $\mu$ with the following probability density
\begin{equation}
P\left(\mu_{ij}\right)=\mathcal{P}\delta\left(\mu_{ij}-\mu\right)+\left(1-\mathcal{P}\right)\delta\left(\mu_{ij}\right).
\label{eq:Pm}
\end{equation}
Thus, either a bond  with a stretch modulus $\mu$ is present with a probability $\mathcal{P}$, or absent with a probability $1-\mathcal{P}$. By varying $\mathcal{P}$ we can tune the connectivity, i.e. the mean number of springs, $z$, which are attached to a crosslink of the network. Here we study nonlinear elasticity of such bond-diluted networks on a triangular lattice in $d=2$ and face centered cubic (FCC) lattice in $d=3$.

%The elastic response of such a network in the linear regime ($\epsilon\rightarrow0$) has been studied extensively~ %\cite{feng1984,Feng85}. Consistent with those studies we find that far from the isostatic point the elastic moduli %depend linearly on the mean coordination number (Fig.~\ref{Fig0}), in agreement with the mean-field theory prediction. %More specifically, the bulk modulus, $B$, vanishes as a power law at the critical point as $B\sim |\Delta z|^f$, where %$\Delta z=z-z_c$ and $f$ is the rigidity exponent; mean field theory predicts $f=1$ consistent with our results and %previous results far from the isostatic point. By contrast, non-mean field behavior, $f\not=1$, is observed close to %the critical point
%\cite{arbabi1993mechanics,head2003nonuniversality,Broedersz2011}.

\begin{figure}[tp]
\begin{centering}
\includegraphics[width=\columnwidth]{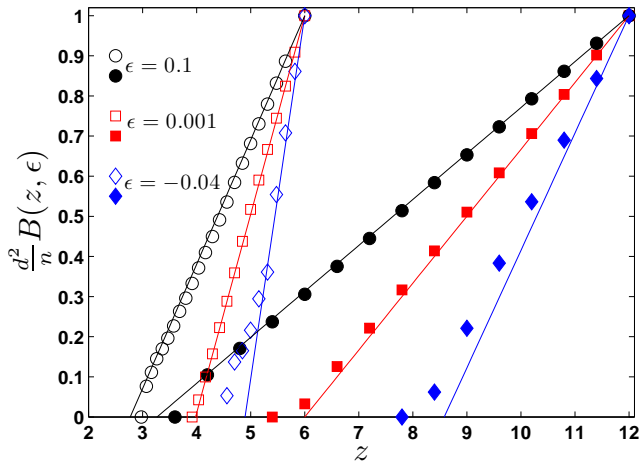}
\par\end{centering}
\caption{(Color online) The nonlinear differential bulk modulus as a function of the coordination number for triangular (open symbols) and FCC (filled symbols) lattices for different strain values. The solid lines are the results of the EM theory (Eq. (\ref{eq:KEM_Diluted}) or, in the explicit form, Eq. (\ref{eq:K_explicit})).}
\label{Fig0}
\end{figure}

The mechanical response of these networks is sensitive to the applied strain. We quantify this network response with a differential bulk modulus, $B\left(z,\epsilon\right)$ as shown in Fig.~\ref{Fig0}. The qualitative behavior of the nonlinear bulk modulus is similar to the behavior of the bulk moduli in the linear regime, but with an isostatic point that shifts continuously to lower coordination numbers, as demonstrated in Fig. \ref{Fig0}. The nonlinear response of the diluted central-force networks can be characterized by a two-dimensional $\left(z,\epsilon\right)$ phase diagram of the differential bulk modulus, as shown schematically in Fig. \ref{phasescheme}. The transition curve, $z_c\left(z\right) $, separates the floppy and the rigid regions. From this it can be understood that a subisostatic diluted regular network with central-force interactions exhibits a strain-stiffening behavior --- from a floppy to a rigid structure --- as a function of applied strain.

\subsection{Mean-field approach}\label{The effective medium approach}
We gain insight in the nonlinear response of random spring networks (see Eq. (\ref{Model})), by developing an effective medium approach for the high strain regime. This EM theory aims to  provide a complete quantitative description of the nonlinear elastic response of a network under an external expansional/compressional strain $\epsilon$. Our approach is a nonlinear extension of the EM approaches used to successfully describe the linear elastic response of diluted lattice based networks \cite{Feng85}, and goes beyond perturbative approaches for networks with small dilution~\cite{tang1988percolation}.
The effective medium theory for the linear elastic response of the
disordered spring networks under small deformation was shown to predict
the location of the critical coordination number and the elastic response
far from it \cite{Feng85,schwartz1985behavior,mao2011coherent}. In other systems with non-central force interactions, such as fiber bending models, the EM theory
succeeded to capture the qualitative elastic behavior of the network \cite{Broedersz2011}.

\begin{figure}[tp]
\begin{centering}
\includegraphics[width=\columnwidth]{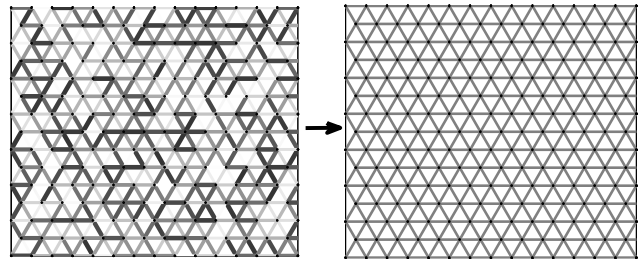}
\par\end{centering}
\caption{Illustration of the EM approach. The network on the left represents the
original system where the disorder in the spring constant is illustrated
by the disorder of the width and the gray level of a bond. The right panel
represents the EM with a regular, non-disordered structure.}
\label{Illustration}
\end{figure}

The nonlinear EM approach developed here is based on a scheme to construct a mapping from the disordered
system onto a perfect lattice system with uniform bond stiffness with the same network topology and strain $\epsilon$, as illustrated in Fig.
\ref{Illustration}. This mapping is realized
by an effective uniform central force interaction, $\mu_{ij}\rightarrow\widetilde{\mu}$.
The effective parameter, $\widetilde{\mu}$, is determined using a
self-consistency requirement: replacing a random bond in the uniform
EM under strain with a bond drawn from the original probability density,
$P\left(\mu_{ij}\right)$, results in a local fluctuation in the deformation
field, which vanishes when averaged over the distribution $P\left(\mu_{ij}\right)$. In addition, we assume that the fluctuations of the deformations are small compared to the distance between crosslinks.
Importantly, the effective parameter $\widetilde{\mu}$
depends on strain, which gives rise to a nonlinear elastic response.

\section{Effective medium theory}
\label{Effective medium theory}
We apply the EM theory method to a network subjected to a global expansion
with a macroscopic isotropic strain $\epsilon$.
The position of a crosslink $i$ is given by $\mathbf{R}_{i}=\mathbf{R}^0_{i}+\mathbf{u}_{i}$,
where $\mathbf{R}^0_{i}$ is the position in the unstrained configuration and $\mathbf{u}_{i}$
is the displacement due to the applied strain.
The affine displacement is defined as $\mathbf{u}_{i}^{\rm aff}-\mathbf{u}_{j}^{\rm aff}=\left(1+\epsilon\right)\mathbf{r}_{ij}$,
where $\mathbf{r}_{ij}$ is the vector from $\mathbf{R}^0_{i}$
to $\mathbf{R}^0_{j}$ in the undeformed reference state. Here we allow for non-affine displacements
\begin{equation}
\mathbf{v}_{i}\equiv\mathbf{u}_{i}-\mathbf{u}_{i}^{\rm aff},
\end{equation}
given by the deviation of the displacement of network node $i$ from its affine value.
However, we assume that the resulting non-affine relative displacements of neighbouring nodes $i$ and $j$ are much
smaller than the corresponding affine displacement:
\begin{equation}
\left|\mathbf{v}_{ij}\right|\equiv\left|\mathbf{v}_i-\mathbf{v}_j\right| \ll \left|\mathbf{u}_{i}^{\rm aff}-\mathbf{u}_{j}^{\rm aff}\right| =  \ell_0\epsilon.
\end{equation}
Thus, we can expand the
Hamiltonian around the affine strain configuration (small $v_{ij}$). Up to second
order in $v_{ij}$ we arrive at~\cite{tang1988percolation}
\begin{equation}
\label{eq:nonlinearmodel}
H=\underset{\left\langle ij\right\rangle }{\sum}\mu_{ij}\left[\frac{1}{2}\epsilon^{2}+\epsilon\mathbf{v}_{ij}\cdot\mathbf{r}_{ij}+\frac{1}{2}\frac{\left(\mathbf{v}_{ij}\cdot\mathbf{r}_{ij}\right)^{2}+\epsilon\mathbf{v}_{ij}^{2}}{1+\epsilon}\right].
\end{equation}
The expansion of the whole network corresponds to the global
constraint
\begin{equation}
\underset{\left\langle ij\right\rangle }{\sum}\mathbf{v}_{ij}=0.
\label{eq:constraint}
\end{equation}

Interestingly, the Hamiltonian for the network under a finite strain bears a resemblance with the Born model, which includes both isotropic and anisotropic pairwise interactions~\cite{born1954dynamical}---as is the case here. However, as was noticed in \cite{tang1988percolation}, the model presented here is formally distinct from the Born model, since here $\mathbf{v}_i$ is the displacement from the affinely deformed state and not from the undeformed configuration as is the case in the Born model. Thus, from our model (Eq.~(\ref{eq:nonlinearmodel})) we see that a finite strain introduces additional interactions that penalize non-affine deformations
with a coupling parameter that is directly proportional to the strain $\epsilon$.

To investigate the nonlinear elastic behavior of the model in Eq.~(\ref{eq:nonlinearmodel}), we set up an effective medium theory.
In the EM approach, we mimic the disordered system by the regular one with
an effective parameter, i.e. $\mu_{ij}\rightarrow\widetilde{\mu}$.
To deform the EM network similarly to the original system with the
same global strain $\epsilon$, one can use a Lagrange multiplier to ensure that the
global constraint (\ref{eq:constraint}) is satisfied. In other words,
the EM network may be globally expanded by applying the force that
assures mechanical equilibrium for the affine, $\mathbf{v}_{ij}=0$, configuration.
Thus, the EM system has the Hamiltonian, given by
\begin{align}
H_{EM} & =\underset{\left\langle ij\right\rangle }{\sum}\widetilde{\mu}\left(\epsilon\right)\left[\frac{1}{2}\epsilon^{2}+\epsilon\mathbf{v}_{ij}\cdot\mathbf{r}_{ij}+\frac{1}{2}\frac{\left(\mathbf{v}_{ij}\cdot\mathbf{r}_{ij}\right)^{2}+\epsilon\mathbf{v}_{ij}^{2}}{1+\epsilon}\right]\label{eq:H_EM}\\
+ & \mathbf{\Lambda}_{ij}\cdot\mathbf{v}_{ij}\nonumber
\end{align}
 where $\mathbf{\Lambda}_{ij}=-\widetilde{\mu}\epsilon\mathbf{r}_{ij}$.
To calculate the effective parameter $\widetilde{\mu}$ we demand
self-consistency of the EM~\cite{Feng85}. The self-consistency requirement in this context can be formulated as follows: the non-affine displacement induced by the replacement of a single
bond in the EM vanishes on average,
\begin{equation}
\left\langle \mathbf{v}_{nm}\right\rangle =0.
\label{eq:SelfConsistence}
\end{equation}
Here, the average is taken over the distribution of the $nm$ bond in the original disordered system, i.e. according to the  probability density $P\left(\mu_{nm}\right)$.
To calculate the displacement $\mathbf{v}_{nm}$ after the replacement
we solve the perturbed EM Hamiltonian that is given by
\begin{align}
H_{EM}+ & \frac{1}{2}\left(\mu_{nm}-\widetilde{\mu}\right)\frac{\left(\mathbf{v}_{nm}\cdot\mathbf{r}_{nm}\right)^{2}+\epsilon\mathbf{v}_{nm}^{2}}{1+\epsilon}\nonumber \\
+ & \mathbf{v}_{nm}\cdot\mathbf{r}_{nm}\epsilon\left(\mu_{nm}-\widetilde{\mu}\right)
\end{align}
 In the configuration that minimizes the energy, the displacement of
the $nm$ bond is given by
\begin{equation}
\mathbf{v}_{nm}=\frac{\mathbf{r}_{nm}\epsilon\left(\mu_{nm}-\widetilde{\mu}\right)}{\mu_{EM}+\mu_{nm}-\widetilde{\mu}},
\label{eq:DisplacementMainText}
\end{equation}
where $\mu_{EM}$ is the displacement of the $nm$ bond in the \emph{unperturbed}
EM network due to a unit force acting along the $nm$
bond. As detailed in Appendix \ref{AppA}, it is given by
\begin{equation}
\mu_{EM}=\frac{\widetilde{\mu}\left(\epsilon\right)}{a\left(\epsilon\right)},
\label{eq:mEMmaintext}
\end{equation}
where $a\left(\epsilon\right)$ is given in Eq. (\ref{eq:mEM}) and may be approximated for a highly coordinated lattice by
\begin{equation}
a\left(\epsilon\right)\approx\frac{2d\left(1+\epsilon\right)}{\mathcal{Z}}\left[1-\frac{\epsilon}{d}\left(\frac{1}{\frac{3}{2+d}+\epsilon}+\frac{d-1}{\frac{1}{2+d}+\epsilon}\right)\right].
\label{eq:ad}
\end{equation}
Given Eqs. (\ref{eq:DisplacementMainText},\ref{eq:mEMmaintext}), the self-consistency Eq. (\ref{eq:SelfConsistence}) leads to the following equation
for the effective parameter\footnote{In the linear regime, Eq. (\ref{eq:mEff_Integral}) reduces to Eq. (9) in Ref. \cite{Feng85}.}
\begin{equation}
\int_{0}^{\infty}\frac{1-\widetilde{\mu}\left(\epsilon\right)/\mu_{ij}}{\frac{1}{a\left(\epsilon\right)}+1-\widetilde{\mu}\left(\epsilon\right)/\mu_{ij}}P\left(\mu_{ij}\right)d\mu_{ij}=0.\label{eq:mEff_Integral}
\end{equation}

Importantly, in contrast to the linear EM, here the effective parameter
$\widetilde{\mu}\left(\epsilon\right)$ cannot be interpreted as the effective spring-constant of the bonds in the EM. However, using the expression for
$\widetilde{\mu}\left(\epsilon\right)$ one can determine the elastic
properties of the original disordered system as follows. Since the equilibrium
configuration of the regular, EM network is given by the affine expansion, $\mathbf{v}_{ij}=0$,
its energy (per bond) is given by
\begin{equation}
E_{EM}\left(\epsilon\right)=H_{EM}\left(\mathbf{v}_{ij}=0\right)=\frac{1}{2}\widetilde{\mu}\left(\epsilon\right)\epsilon^{2}.
\label{eq:EEM}
\end{equation}
The last expression may be interpreted as an approximation for the
original system's energy up to correction terms. These terms appear since the energy is defined as
\begin{equation}
E_{EM}\left(\epsilon\right)=\frac{d^2}{n}\int_{0}^{\epsilon}\int_{0}^{\epsilon'}B_{EM}\left(\epsilon''\right)d\epsilon''\epsilon'
\label{eq:EvsK}
\end{equation}
or
\begin{equation}
E_{EM}\left(\epsilon\right)=\frac{\left(1+\epsilon\right)^d}{n}\int_{0}^{\epsilon}\Pi_{EM}\left(\epsilon'\right)\epsilon'.
\label{eq:EvsSig}
\end{equation}
Thus, for $z<z_0$ it includes the integration in the floppy phase, where the EM theory breaks
down and predicts non-physical elastic properties. To calculate the correction terms we calculate first the floppy and rigid phases separation curve and then subtract from the expression (\ref{eq:EEM}) the integration in the floppy phase. For a given mean coordination number the transition strain value, $\epsilon_{c}\left(z\right)$, may be found as follows. If the transition is first order, one requires that the energy and its first derivative vanish at the transition point. By contrast, if the transition is second order, one requires that the energy and its first and second derivatives vanish at the transition point. These two possible assumptions about the transition order result in different transition curves $\left(z_{c},\epsilon_c\right)$. We define them as $\left(z_{c_1},\epsilon_{c_1}\right)$ for the first order transition case and $\left(z_{c_2},\epsilon_{c_2}\right)$ for the second order transition case.
Since the order of the transition cannot be deduced from the EM theory, described here, we will analyze both options. In Sec. \ref{Diluted regular networks} we calculate both transition curves for the particular example of the diluted regular networks and compare them to the numerical results.

The nonlinear bulk modulus, defined in Eq. (\ref{K_def}), does not depend on the transition order and may be approximated by the EM approach using Eqs.
(\ref{eq:EEM}) and (\ref{eq:EvsK}). It is given by
\begin{equation}
B_{EM}\left(\epsilon\right)=\frac{n}{d^2}\frac{\partial^{2}}{\partial\epsilon^{2}}\left[\widetilde{\mu}\left(\epsilon\right)\frac{\epsilon^{2}}{2}\right],
\label{eq:KEM_general}
\end{equation}
 where $\widetilde{\mu}\left(\epsilon\right)$ is given in Eq. (\ref{eq:mEM}) and
is approximated in Eq. (\ref{eq:ad}).
In the limit of small strain, given by $\epsilon\rightarrow 0$, the mean-field rigidity
threshold of the network, $z_{c}\left(\epsilon\rightarrow 0\right)=2d$, also does not depend on the transition order.
This corresponds to the mean-field isostatic point \cite{Maxwell1864,Feng85}.
In the limit of large strain, $\epsilon\rightarrow\infty$, the mean-field rigidity
threshold of the network does not depend on the transition order and is $z_{c}\left(\epsilon\rightarrow\infty\right)=2$.

The proper energy and the pressure in the system can be obtained from $B_{EM}\left(\epsilon\right)$ by integration from $\epsilon_c$
\begin{equation}
E_{EM}\left(\epsilon\right)=\frac{d^2}{n}\int_{\epsilon_{c}}^{\epsilon}\int_{\epsilon_{c}}^{\epsilon'}B_{EM}\left(\epsilon''\right)d\epsilon''d\epsilon'\end{equation}
 and
\begin{equation}
\Pi_{EM}\left(\epsilon\right)=\frac{d^2}{\left(1+\epsilon\right)^d}\int_{\epsilon_{c}}^{\epsilon}B_{EM}\left(\epsilon'\right)d\epsilon',
\end{equation}
where $\epsilon_c$ is defined as $\epsilon_{c_1}$ or $\epsilon_{c_2}$ for the first and second order transition assumptions, respectively.

The approach presented in this section allows one to calculate the elastic
parameters of a system with a given topology and elastic constant
distribution in the nonlinear elastic regime. The nonlinear differential bulk
modulus is given by Eq. (\ref{eq:KEM_general}) while Eq. (\ref{eq:mEff_Integral}) determines the effective parameter
$\widetilde{\mu}\left(\epsilon\right)$. Eq. (\ref{eq:mEff_Integral}) may be solved numerically for any realization of the spring constant probability density, $P\left(\mu_{ij}\right)$. In Sec. \ref{Diluted regular networks} we demonstrate the presented method using the particular example of diluted regular
networks when Eq. (\ref{eq:mEff_Integral}) can be easily solved analytically.

\subsection{Mapping fluctuations to random forces}
\label{Mapping}
Within the EM approach, desribed above, a deviation of the spring constant of a given bond from the EM spring constant is described as an additional force dipole that act on this bond. Therefore, the disorder of the spring constant may be mapped to the disorder of the force dipoles which act on the regular EM in the EM theory approximation.
More specifically, the replacement of the EM spring between nodes $i$ and
$j$ by the spring with the elastic constant $\mu_{ij}$ is equivalent
to the force
\begin{equation}
\mathbf{f}_{n}^{ij}=\epsilon\frac{\widetilde{\mu}-\mu_{ij}}{\frac{\widetilde{\mu}}{a\left(\epsilon\right)}-\widetilde{\mu}+\mu_{ij}}\frac{\widetilde{\mu}}{a\left(\epsilon\right)}\left(\delta_{i,n}-\delta_{j,n}\right)\mathbf{r}_{ij}
\label{eq:force}
\end{equation}
acting along the bond. Due to the self-consistence requirement (\ref{eq:SelfConsistence})
the average force is zero,
\begin{equation}
\left\langle \mathbf{f}_{n}^{ij}\right\rangle =0
\label{ForceVanishes}
\end{equation}
and, since we assumed that there is no correlation between spring
constants on distinct bonds, the associated forces are also uncorrelated:
\begin{equation}
\left\langle \mathbf{f}_{n}^{ij}\cdot\mathbf{f}_{n}^{i'j'}\right\rangle =\delta_{i,i'}\delta_{j,j'}\left\langle \left(\mathbf{f}_{n}^{ij}\right)^{2}\right\rangle.
\label{ForcesUncorrelated}
\end{equation}
The obtained regular lattice (EM) with random forces is equivalent
to \emph{model B} in Ref. \cite{didonna2005} (in contrast to the original system which is defined as \emph{model A} in Ref. \cite{didonna2005}) and may thus be treated similarly. In particular, the non-affinity correlation function of the strain field may be evaluated within this model. The described mapping has the same level of approximation as the EM approximation. However, this mapping completes the EM theory in the sense that it allows to calculate all the correlation functions. Below we provide two examples of the usefulness of this mapping by calculating two important one-point correlation functions: the average non-affine displacements and its nonlinear, differential analog.
\subsection{Correlation functions}
\label{CorrFunctions}
It is known that the rigidity percolation transition in the small-strain
limit is accompanied by highly non-affine network deformations \cite{jacobs1996,wyart2008,Broedersz2011}.
As we show in this work, the non-affine deformation field is responsible for the strain stiffening behavior of the disordered spring networks. The mapping, described in Sec. \ref{Mapping}, allows us to calculate any correlation function of the displacement field.
\subsubsection{Non-affinity parameter}
Several methods have been proposed to quantify the deviation from
a uniform (affine) strain field \cite{head2003distinct,onck2005alternative,didonna2005,liu2007}.
A useful parameter for the non-affinity
characterization is the average deviation from the affine configuration,
\begin{equation}
\Gamma=\frac{1}{\epsilon^{2}}\left\langle \mathbf{v}_{n}^{2}\right\rangle _{n},
\end{equation}
 where $\left\langle \cdots\right\rangle _{n}$ denotes the average
over all vertices.
Below we calculate $\Gamma$ using the EM approach including the mapping described in Sec. \ref{Mapping}.

The Fourier transform of the force (\ref{eq:force}) is given by
\begin{equation}
\mathbf{f}^{ij}\left(\mathbf{k}\right)=\epsilon\frac{\widetilde{\mu}-\mu_{ij}}{\frac{\widetilde{\mu}}{a\left(\epsilon\right)}-\widetilde{\mu}+\mu_{ij}}\frac{\widetilde{\mu}}{a\left(\epsilon\right)}\left(e^{i\mathbf{k}\cdot\mathbf{R}_{i}}-e^{i\mathbf{k}\cdot\mathbf{R}_{j}}\right)\mathbf{r}_{ij}.
\end{equation}
Thus, the Fourier transform of the displacement field from the affine
configuration due to this force is given by
\begin{eqnarray}
\mathbf{v}\left(\mathbf{k}\right) & = & -\underset{\left\langle ij\right\rangle }{\sum}D^{-1}\left(\mathbf{k}\right)\mathbf{f}^{ij}\left(\mathbf{k}\right)\nonumber \\
 & = & \frac{\epsilon\left(1+\epsilon\right)}{a\left(\epsilon\right)}\underset{\left\langle ij\right\rangle }{\sum}\frac{\frac{\widetilde{\mu}-\mu_{ij}}{\frac{\widetilde{\mu}}{a\left(\epsilon\right)}-\widetilde{\mu}+\mu_{ij}}\left(e^{i\mathbf{k}\cdot\mathbf{R}_{i}}-e^{i\mathbf{k}\cdot\mathbf{R}_{j}}\right)\mathbf{r}_{ij}}{\underset{\mathbf{r}}{\sum}\left(\mathbf{r}\otimes\mathbf{r}+\epsilon\mathbb{I}\right)\left(1-e^{i\mathbf{k}\cdot\mathbf{r}}\right)},
 \label{DispToForce}
 \end{eqnarray}
 or in real space
 \begin{equation}
\mathbf{v}_{n}=\frac{\epsilon\left(1+\epsilon\right)}{Na\left(\epsilon\right)}\underset{\left\langle ij\right\rangle ,\mathbf{k}}{\sum}\frac{\frac{\widetilde{\mu}-\mu_{ij}}{\frac{\widetilde{\mu}}{a\left(\epsilon\right)}-\widetilde{\mu}+\mu_{ij}}\left(e^{i\mathbf{k}\cdot\mathbf{R}_{i}}-e^{i\mathbf{k}\cdot\mathbf{R}_{j}}\right)e^{-i\mathbf{k}\cdot\mathbf{R}_{n}}\mathbf{r}_{ij}}{\underset{\mathbf{r}}{\sum}\left(\mathbf{r}\otimes\mathbf{r}+\epsilon\mathbb{I}\right)\left(1-e^{i\mathbf{k}\cdot\mathbf{r}}\right)}.
\end{equation}
 Since the forces are independent and identically distributed random variables,
the variance of the displacement from the affine configuration of
every network crosslink is given by
\begin{align}
\left\langle \mathbf{v}_{n}^{2}\right\rangle _{n}& = \frac{1}{2}\left[\frac{\epsilon\left(1+\epsilon\right)}{a\left(\epsilon\right)}\right]^{2}\left\langle \left(\frac{\widetilde{\mu}-\mu_{ij}}{\frac{\widetilde{\mu}}{a\left(\epsilon\right)}-\widetilde{\mu}+\mu_{ij}}\right)^{2}\right\rangle \nonumber \\
&\times \frac{1}{N}\underset{\mathbf{r},\mathbf{k}}{\sum}\left|\frac{1-e^{i\mathbf{k}\cdot\mathbf{r}}}{\underset{\mathbf{r}}{\sum}\left(\mathbf{r}
\otimes\mathbf{r}+\epsilon\mathbb{I}\right)\left(1-e^{i\mathbf{k}\cdot\mathbf{r}}\right)}\mathbf{r}\right|^{2}.
\end{align}
The same approximation of highly coordinated lattice that was used to derive Eq. (\ref{eq:ad_App})
now gives
\begin{eqnarray}
\left\langle \mathbf{v}_{n}^{2}\right\rangle _{n}& \simeq \frac{d\epsilon^{2}\left(1+\epsilon\right)^{2}}{2\mathcal{Z}a^{2}\left(\epsilon\right)}\left\langle \left(\frac{\widetilde{\mu}-\mu_{ij}}{\frac{\widetilde{\mu}}{a\left(\epsilon\right)}-\widetilde{\mu}+\mu_{ij}}\right)^{2}\right\rangle \nonumber \\
 & \times \left[\frac{\frac{3}{2+d}}{\left(\frac{3}{2+d}+\epsilon\right)^{2}}+\frac{\frac{d-1}{2+d}}{\left(\frac{1}{2+d}+
 \epsilon\right)^{2}}\right]\frac{\underset{\mathbf{k}}{\sum}\frac{1}{k^{2}}}{N}.
 \end{eqnarray}
The sum over $\mathbf{k}$ may be estimated for any dimension
\begin{equation}
\frac{1}{N}\underset{\mathbf{k}}{\sum}\frac{1}{k^{2}}=A_{d}\ell_{0}^{2}f_{d}\left(N\right),
\end{equation}
where
\begin{equation}
f_{d}\left(N\right)=\begin{cases}
\frac{d}{2-d}N^{\frac{2}{d}-1} & d<2\\
\ln N & d=2\\
\frac{d}{d-2} & d>2\end{cases}
\end{equation}
and
\begin{equation}
A_{d}=\frac{\frac{1}{N}\underset{\mathbf{k}}{\sum}\frac{1}{k^{2}}}{\frac{1}{\int_{1/L}^{1/\ell_{0}}k^{d-1}dk}\int_{1/L}^{1/\ell_{0}}\frac{k^{d-1}}{k^{2}}dk}=\mathcal{O}\left(1\right)
\end{equation}
is a dimensionless parameter which depends weakly on the lattice geometry
(for instance, $A_{2}\simeq0.36$ for the triangular lattice). As defined
above, $\ell_{0}$ is the rest length of a bond (throughout this paper
this quantity was set to unity, but is shown here explicitly to emphasize
the cut-off of the integral in Fourier space and the units of the
non-affine parameter) and $L=\ell_{0}N^{1/d}$ is the size of the unstrained
network. In sum, the non-affinity parameter is given by
\begin{widetext}
\begin{eqnarray}
\Gamma\left(\epsilon\right)\simeq & A_{d}\ell_{0}^{2} & \frac{d\left(1+\epsilon\right)^{2}}{2\mathcal{Z}a^{2}\left(\epsilon\right)}\left\langle \left(\frac{\widetilde{\mu}-\mu_{ij}}{\frac{\widetilde{\mu}}{a\left(\epsilon\right)}-\widetilde{\mu}+\mu_{ij}}\right)^{2}\right\rangle \left[\frac{\frac{3}{2+d}}{\left(\frac{3}{2+d}+\epsilon\right)^{2}}+\frac{\frac{d-1}{2+d}}{\left(\frac{1}{2+d}+\epsilon\right)^{2}}\right]f_{d}\left(N\right)
\label{Gamma}
\end{eqnarray}
\end{widetext}

\subsubsection{Differential non-affinity parameter}
In the nonlinear regime the more interesting quantity is the differential
non-affinity fluctuation defined as
\begin{align}
&\delta\Gamma\left(\epsilon\right)= \left\langle \left[\underset{\Delta\epsilon\rightarrow0}{\lim}\frac{\mathbf{v}_{n}
\left(\epsilon+\Delta\epsilon\right)-\frac{1+\epsilon+\Delta\epsilon}{1+\epsilon}\mathbf{v}_{n}
\left(\epsilon\right)}{\Delta\epsilon}\right]^{2}\right\rangle _{n} \nonumber \\
= & \left\langle \left(\frac{d\mathbf{v}_{n}\left(\epsilon\right)}{d\epsilon}-\frac{\mathbf{v}_{n}\left(\epsilon\right)}{1+\epsilon}\right)^{2}\right\rangle _{n} \nonumber \\
= & \left\langle \left(\frac{d\mathbf{v}_{n}\left(\epsilon\right)}{d\epsilon}\right)^{2}\right\rangle _{n}-\frac{1}{1+\epsilon}\frac{d\left[\epsilon\Gamma\left(\epsilon\right)\right]}{d\epsilon}+\left(\frac{\epsilon}{1+\epsilon}\right)^{2}\Gamma\left(\epsilon\right).
\label{DiffNonAff}
\end{align}
The first term in the last expression is calculated in Appendix \ref{CalcDiffNonAffinity} while the last two may be easily deduced from Eq. (\ref{Gamma}).
\subsection{Ginzburg criterion}
\label{Ginzburg}
Although the mean-field approach is not a controlled approximation and has no small parameter, one can check for self-consistency of the assumption of small fluctuations \cite{ginzburg1960}. In our case we assume small relative deviations of the EM from the affine strain field (see Eq. (\ref{eq:nonlinearmodel})):
\begin{equation}
\frac{\left\langle\mathbf{v}^2_{nm}\right\rangle_{\left\langle nm \right\rangle}}{\epsilon^2} \ll 1,
\end{equation}
where $\left\langle...\right\rangle_{\left\langle nm \right\rangle}$ is the average over all connected nodes of the EM network.
Therefore, it is instructive to analyze the behavior of the two point, nearest neighbour correlation function. Using the same mapping to the random forces model as above one gets
\begin{equation}
\Gamma_{NN} = \frac{\left\langle\mathbf{v}^2_{nm}\right\rangle_{\left\langle nm \right\rangle}}{\epsilon^2} = \frac{\Gamma}{A_d f_d}\frac{B_d}{d^2},
\label{Ginzburg1}
\end{equation}
where
\begin{equation}
B_{d}=\frac{1}{N}\underset{\mathbf{k}}{\sum}k^{2}\simeq\frac{\int_{1/L}^{1/\ell_{0}}k^{d-1}k^{2}dk}{\int_{1/L}^{1/\ell_{0}}k^{d-1}dk}.
\end{equation}
The nonlinear, differential version of $\Gamma_{NN}$, defined as
\begin{equation}
\delta\Gamma_{NN}=  \left\langle \left[\underset{\Delta\epsilon\rightarrow0}{\lim}\frac{\mathbf{v}_{nm}\left(\epsilon+\Delta\epsilon\right)-\frac{1+\epsilon+\Delta\epsilon}{1+\epsilon}\mathbf{v}_{nm}\left(\epsilon\right)}{\Delta\epsilon}\right]^{2}\right\rangle _{\left<nm\right>},
\end{equation}
is given by
\begin{equation}
\delta\Gamma_{NN} = \frac{\delta\Gamma}{A_d f_d}B_d.
\label{Ginzburg2}
\end{equation}

In Sec. \ref{DilutedNetworksFluctuations} we analyze the non-affinity parameters for the
particular example of the diluted regular networks and compare the
analytical results with the numerical simulations.

\section{Diluted regular networks} \label{Diluted regular networks}
A significant understanding of different physical phenomena in disordered
systems, including percolation \cite{HavlinBook2000,sahimi2003Book} and the
elastic behavior of amorphous materials \cite{sahimi2003Book} was achieved by modelling the topological disorder by a random dilution of a regular
structure. Motivated by this we demonstrate the mean-field solution presented above using the particular example of bond-diluted regular networks. The probability density for the spring constants for such a network is given in Eq.~(\ref{eq:Pm}).
Networks of this kind are referred to as diluted spring networks or the central-force elastic percolation
model. The linear elastic response of this model has been extensively studied \cite{feng1984,Feng85}. Here we show how these results generalize for large strain values.
Before presenting the full mean-field solution for these networks, below we briefly sum up the known relevant results in the small strain regime and discuss the infinite strain limit expectations from the nonlinear EM theory.

\subsection{Zero strain limit}\label{Zero strain limit}
The average coordination number for bond diluted networks is defined as
\begin{equation}
z=\left\langle \underset{j}{\sum}\left(1-\delta_{\mu_{ij},0}\right)\right\rangle _{i}=\mathcal{Z}\mathcal{P},
\label{eq:CoordiantionNumber}
\end{equation}
 where $\left\langle \cdots\right\rangle _{i}$ denotes an average
over all network vertices and $\mathcal{Z}$ is the coordination number
given that all existing springs have a non-zero spring constant. Below
the so-called isostatic threshold of the average coordination number, $z_0 \equiv z_{c}\left(\epsilon\rightarrow0\right)$, the network is
floppy \cite{Maxwell1864,Alexander1998}.

For the unstressed reference state and zero strain limit, Maxwell introduced
a mean-field counting argument for this threshold coordination number
at which the number of degrees of freedom and the number of constraints
due to the central force interactions are equal. This yields an EM
approximation for the isostatic coordination number
\begin{equation}
z_0=2d.
\label{eq:Maxwell}
\end{equation}
It was conjectured (see Refs. in \cite{Feng85}) that the
bulk modulus of the diluted network in the zero strain limit can be expressed in term of $z_0$ as
\begin{equation}
B\left(\epsilon\rightarrow0\right)=\mu\frac{n}{d^2}\frac{z-z_0}{\mathcal{Z}-z_0}.
\label{eq:LinearBulkModulus}
\end{equation}

Eqs. (\ref{eq:Maxwell},\ref{eq:LinearBulkModulus}) were derived
\cite{Feng85} using the EM theory and were shown to predict well
the location of the isostatic point and the elasticity of a
diluted network far from its isostatic point.

\subsection{Infinite strain limit}
Before we turn to the full problem with arbitrary strain values, it
is instructive to discuss our expectations in
the infinite strain limit. In this limit, the rigidity
threshold (isostatic point) can be expected to approach the conductivity threshold,
denoted here by $z_{cond}$. By analogy with the behavior at small strains Eq. (\ref{eq:LinearBulkModulus}), we anticipate (and derive this
result below) that in the large strain limit the nonlinear bulk modulus
is equal to
\begin{equation}
B\left(\epsilon\rightarrow\infty\right)=\mu\frac{n}{d^2}\frac{z-z_{c}\left(\epsilon\rightarrow\infty\right)}{\mathcal{Z}-z_{c}\left(\epsilon\rightarrow\infty\right)}=\mu\frac{n}{d^2}\frac{z-z_{cond}}{\mathcal{Z}-z_{cond}}.\label{eq:InfStrainBulkMod}
\end{equation}
 The mean-field calculation \cite{kirkpatrick1973} of the conductivity
percolation and our calculation below in the infinite strain limit
both suggest $z_{c}\left(\epsilon\rightarrow\infty\right)=2$. We expect a deviation of the EM theory from the numerical calculation
in the infinite strain limit close to the conductivity percolation
point due to the failure of the mean-field approach to predict the precise value of the conductivity threshold.

\subsection{Full solution}
\begin{figure*}
\begin{centering}
\includegraphics[clip,width=\textwidth]{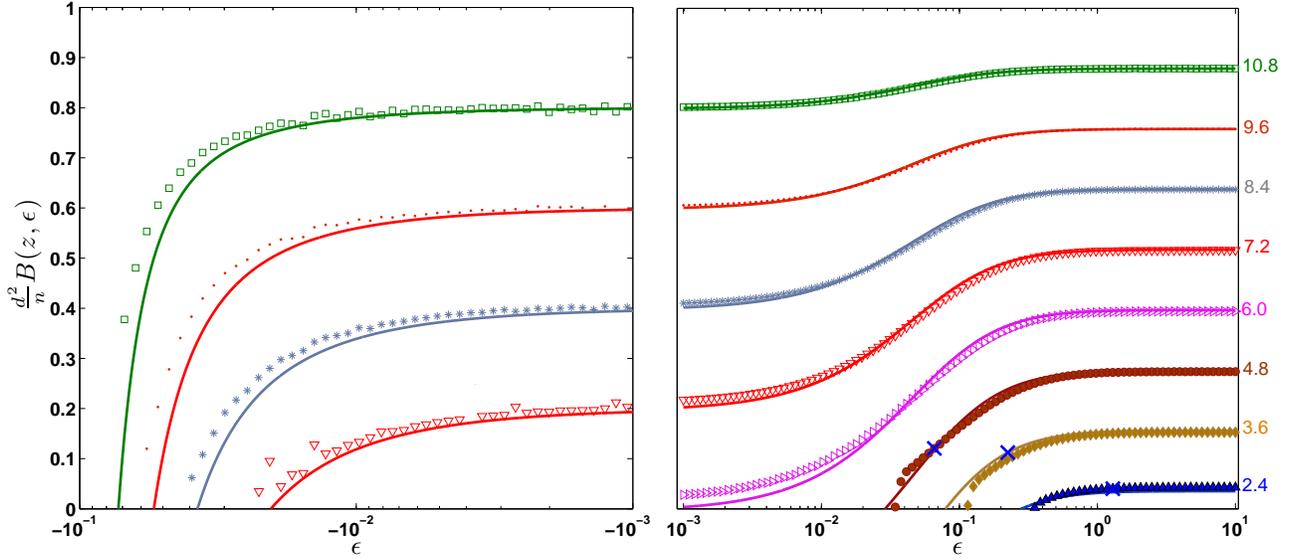}
\par\end{centering}
\caption{The nonlinear bulk modulus for the diluted FCC lattice as a function
of the applied strain for different values of the average coordination
number shown as labels next to the curves. The numerical data is depicted as symbols and the nonlinear
EM theory predictions are shown as solid lines. Big crosses represent the location of the first order transition as predicted by Eq. (\ref{eq:zc_1})}
\label{Fig1}
\end{figure*}
\begin{figure}
\begin{centering}
\includegraphics[clip,width=\columnwidth]{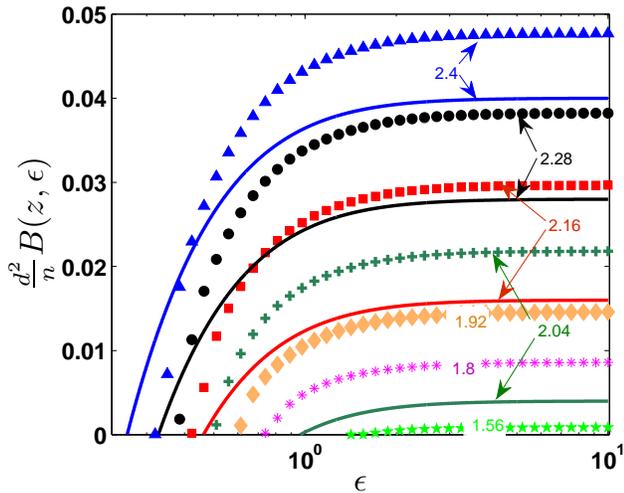}
\par\end{centering}
\caption{The nonlinear bulk modulus for the diluted FCC lattice as a function
of the applied strain for different values of the average coordination
number shown as labels next to the curves. The numerical data is depicted as symbols and the nonlinear
EM theory predictions are shown as solid lines. For $z<2$ the EM theory predicts zero bulk modulus for any strain.}
\label{Fig1b}
\end{figure}
\begin{figure*}
\begin{centering}
\includegraphics[width=\textwidth]{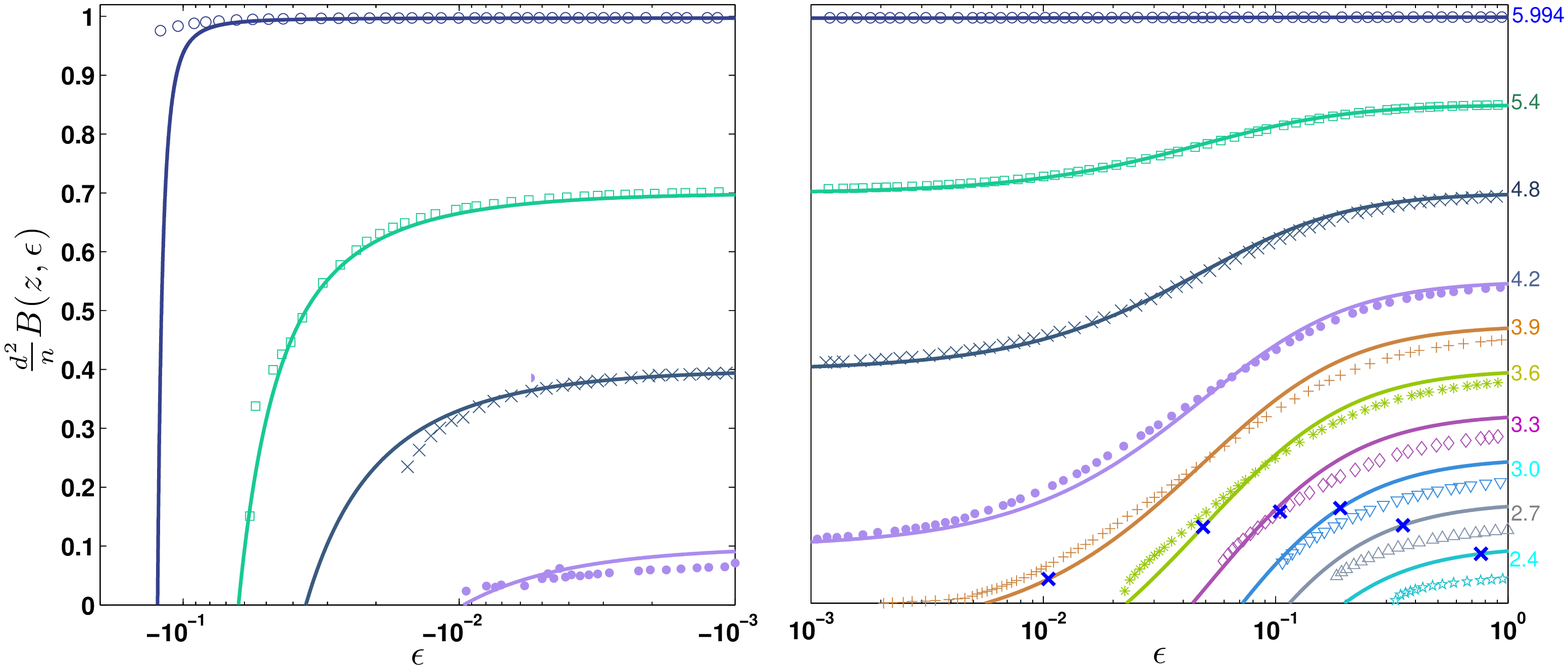}
\par\end{centering}
\caption{The nonlinear bulk modulus for the diluted triangular lattice as a function
of the applied strain for different values of the average coordination
number  shown as labels next to the curves. The numerical data is depicted as symbols and the nonlinear
EM theory predictions are shown as solid lines. Big crosses represent the location of the first order transition as predicted by Eq. (\ref{eq:zc_1})}
\label{Fig2}
\end{figure*}
Using Eq. (\ref{eq:Pm}), the solution for the self consistent equation
(\ref{eq:mEff_Integral}) is given by
\begin{equation}
\widetilde{\mu}\left(\epsilon\right)=\mu\frac{z-a\left(\epsilon\right)\mathcal{Z}}{\mathcal{Z}-a\left(\epsilon\right)\mathcal{Z}},
\label{eq:mtilda}
\end{equation}
 where $a\left(\epsilon\right)$ is given in Eq. (\ref{eq:mEM}) and
is approximated in Eq. (\ref{eq:ad}). The nonlinear differential bulk
modulus of the original system may be approximated by the EM approach
using Eqs. (\ref{eq:KEM_general}) and (\ref{eq:mtilda}) and is given
by
\begin{equation}
B_{EM}\left(z,\epsilon\right)=\begin{cases}
\mu\frac{n}{d^{2}}\frac{\partial^{2}}{\partial\epsilon^{2}}\left[\frac{z-a\left(\epsilon\right)\mathcal{Z}}{\mathcal{Z}-a\left(\epsilon\right)\mathcal{Z}}\frac{\epsilon^{2}}{2}\right] & \epsilon\geq\epsilon_{c}\\
0 & \epsilon<\epsilon_{c}\end{cases},
\label{eq:KEM_Diluted}
\end{equation}
where $\epsilon_c$ is defined as $\epsilon_{c_1}$ or $\epsilon_{c_2}$ for the first and second order transition assumptions, respectively.
In Appendix \ref{AppB} we present the explicit
result for the nonlinear differential bulk modulus and the transition curves (for both assumptions for the transition order), based on Eq. (\ref{eq:KEM_Diluted}).

A comparison between this analytic prediction and the numerical results is shown in Figs. \ref{Fig1},\ref{Fig1b},\ref{Fig2}. Below the conductivity percolation threshold, $z<z_{cond}$, a network, does not resist deformation for any strain and $B\left(\epsilon\right)=0$ due to the absence of an infinite connected cluster. By contrast, when the coordination number is in the range $z_{cond}<z<z_0$, a network only develops a non-zero differential bulk modulus for positive strains above a threshold $\epsilon_c\left(z\right)$. For superisostatic coordination numbers, $z>z_0$, the differential bulk modulus is larger than zero in the small strain limit; $B$ increases with $\epsilon$ until it reaches a plateau of the large strain limit (see right panels in Figs. \ref{Fig1},\ref{Fig2}).

\begin{figure}
\begin{centering}
\includegraphics[width=\columnwidth]{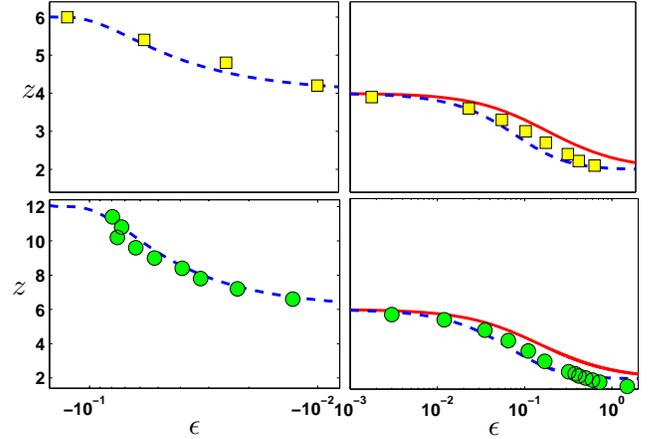}
\par\end{centering}
\caption{The phase diagram for triangular (upper panels) and FCC (bottom panels) lattices. The numerical data for the transition points is depicted as symbols and the nonlinear
EM theory predictions are shown as solid lines for the first order transition given by Eq. (\ref{eq:zc_1}) and dashed lines for the second order transition given by Eq. (\ref{eq:zc_2}). The curves separate the floppy (below) from the rigid (above) phases.}
\label{PhaseDiagram}
\end{figure}

In contrast to the positive strains, for a negative values of the strain, the modulus $B\left(z,\epsilon\right)$ of superisostatic networks decreases with $\left|\epsilon\right|$ until it vanishes below a threshold, predicted by the nonlinear EM theory in Eq. (\ref{eq:zc_2}) (see left plots in Figs. \ref{Fig1},\ref{Fig2},\ref{PhaseDiagram}). This collapse was observed for perfect two-dimensional triangular network in Monte-Carlo simulations at low temperatures \cite{discher1997phase,wintz1997mesh}. Here we show that reduction of the mean coordination number shifts this collapse towards smaller values of $\left|\epsilon\right|$.

The agreement with the numerical data is good far from the transition point. The bulk modulus in the infinite strain limit is given by
\begin{equation}
B_{EM}\left(\epsilon\rightarrow\infty\right)=\mu\frac{n}{d^2}\frac{z-2}{\mathcal{Z}-2},
\end{equation}
such that the transition average coordination number approaches the conductivity threshold, $z_{c}\left(\epsilon\rightarrow\infty\right)=2$.
However, the mean-field
prediction for the conductivity threshold deviates from the numerical result \cite{shante1971,kirkpatrick1973}.
This may account for the discrepancy between the nonlinear EM theory prediction
and the simulation results close to the conductivity percolation threshold
in the large strain regime (see Fig. \ref{Fig1b} and large strain
values for $d=3$ in Fig. \ref{PhaseDiagram}). In fact, we find that for the FCC lattice the rigidity percolation in the large strain limit occurs at $z_c\left(\epsilon\rightarrow\infty\right)=z_{cond}=1.5\pm0.3$. This result is consistent with both the empirical law for the conductivity threshold $z_{cond}\simeq\frac{d}{d-1}$ \cite{shante1971} and with the numerical result of the FCC lattice conductivity threshold $z_{cond}\simeq1.442$ \cite{lorenz1998precise}.

The results discussed above are summarized in a phase diagram shown in Fig.~\ref{PhaseDiagram}.
The curves indicate the transition connectivity number between rigid and floppy phases as a function of applied strain. The explicit formulas may be found in Appendix
\ref{AppB} for both assumptions about the transition order (see Eqs. (\ref{eq:zc_1}) and (\ref{eq:zc_2})). The strain dependence of the isostatic point we find numerically is reasonably well
described by the nonlinear EM theory, as shown in Fig. \ref{PhaseDiagram}. For the negative strain values at the transition point only the differential bulk modulus vanishes but not the stress and the elastic energy; this unambiguously corresponds to a second order transition.
\subsection{Non-affine fluctuations}
\label{DilutedNetworksFluctuations}
\begin{figure*}
\begin{centering}
\includegraphics[width=\textwidth]{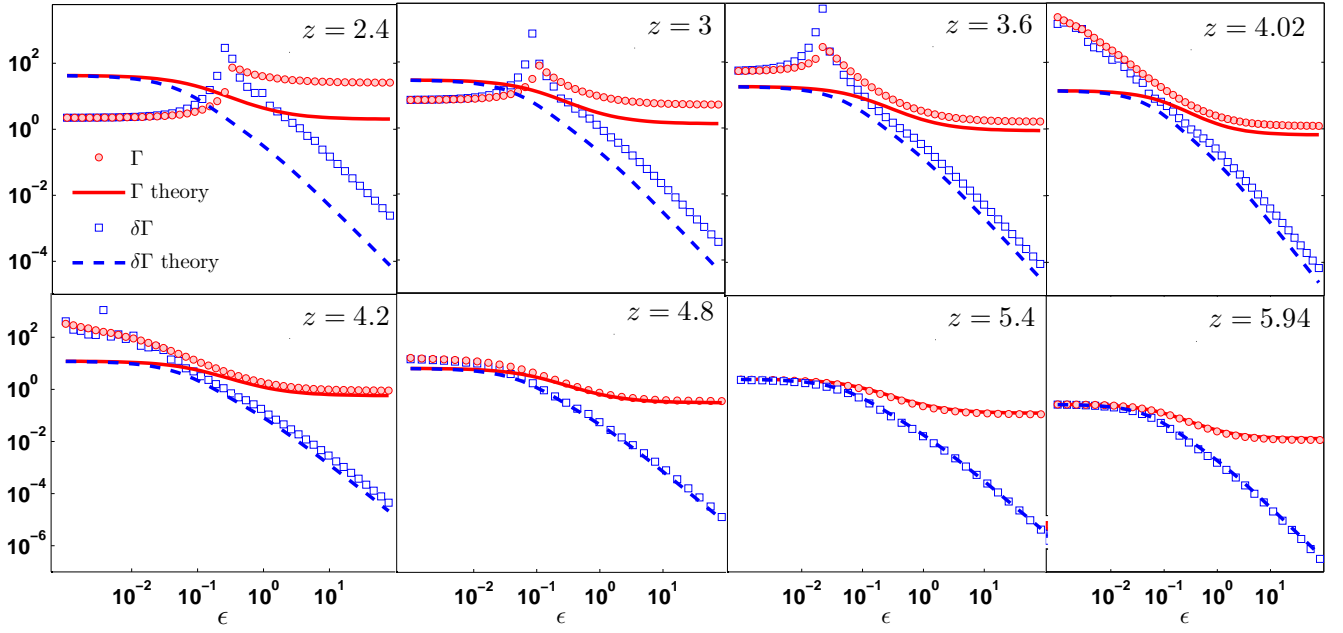}
\par\end{centering}
\caption{Non-affinity one-point correlation functions $\Gamma$ (red filled circles) and $\delta\Gamma$ (blue empty squares) as a function of the applied strain, $\epsilon$ for different values of the mean coordination number, $z$ (see upper right corner of every plot) on the diluted triangular lattice of size $140 \times 140$. The lines represent the theoretical prediction for $\Gamma$ (red solid line) and $\delta\Gamma$ (blue dashed line).}
\label{NonAffineFigure}
\end{figure*}

Here we demonstrate the method presented in Secs. \ref{Mapping} and \ref{CorrFunctions} and analyze the non-affine fluctuations for the particular case of the diluted regular lattices. Using Eqs. (\ref{Gamma}), (\ref{DiffNonAff}) and the expression for the effective parameter, $\widetilde{\mu}\left(\epsilon\right)$, Eq. (\ref{eq:mtilda}), one obtains the expressions for the non-affinity parameters $\Gamma$ and $\delta\Gamma$. A comparison between the analytical formula and the numerical calculation is shown in Fig. \ref{NonAffineFigure}. For superisostatic networks the numerical results agree well with the nonlinear EM theory predictions. However, on the transition curve, the non-affinity parameter appears to diverge; this divergence is not captured by the nonlinear EM theory. Further insight in this discrepancy between the EM theory and numerical results over a range of parameters (including the divergences) is gained by analysing two-point correlation functions.

\begin{figure*}
\begin{centering}
\includegraphics[width=\textwidth]{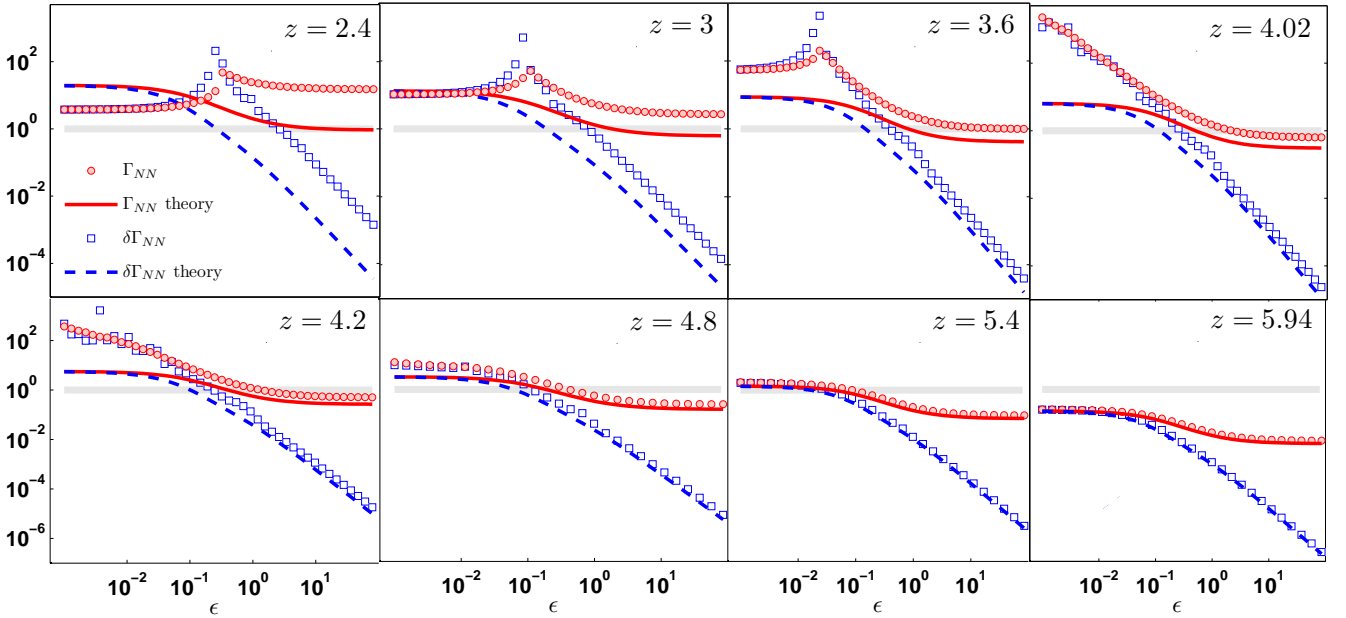}
\par\end{centering}
\caption{Non-affinity two-point correlation functions $\Gamma_{NN}$ (red filled circles) and $\delta\Gamma_{NN}$ (blue empty squares) as a function of the applied strain, $\epsilon$ for different values of the mean coordination number, $z$ (see upper right corner of every plot) on the diluted triangular lattice. The lines represent the theoretical prediction for $\Gamma_{NN}$ (red solid line) and $\delta\Gamma_{NN}$ (blue dashed line). The thick gray line indicate the value of $1$ to compare with $\Gamma_{NN}$ for the Ginzburg criterion.}
\label{NonAffineBondFigure}
\end{figure*}

As discussed in Sec. \ref{Ginzburg}, the mean-field assumption of small fluctuations can be determined self-consistently using the two point, nearest neighbour correlation function defined in Eq. (\ref{Ginzburg1}).
Based on Eq. (\ref{Ginzburg2}) one may calculate the non-affine
fluctuations, $\Gamma_{NN}$, and its differential analog, $\delta\Gamma_{NN}$, for the expanded diluted regular networks. The comparison between the theoretical calculation of $\Gamma_{NN}$ and $\delta\Gamma_{NN}$, including the numerical results is shown in Fig. \ref{NonAffineBondFigure}. The agreement between the theoretical prediction and the numerical data is good when the Ginzburg criterion is satisfied, i.e. $\Gamma_{NN}\ll1$. Clearly, close to the transition curve, where the non-affinity parameters diverge, one can expect the EM theory to fail since the Ginzburg criterion is strongly violated.
Note, however, that the theoretical prediction for the bulk elastic properties appears to be reasonable even when the Ginzburg criterion is not satisfied (see Figs. \ref{Fig2} and \ref{NonAffineBondFigure} for small values of $z$).

\subsection{Additional weak interactions: fiber bending}\label{Additional interactions}
\begin{figure*}
\begin{centering}
\includegraphics[clip,width=\textwidth]{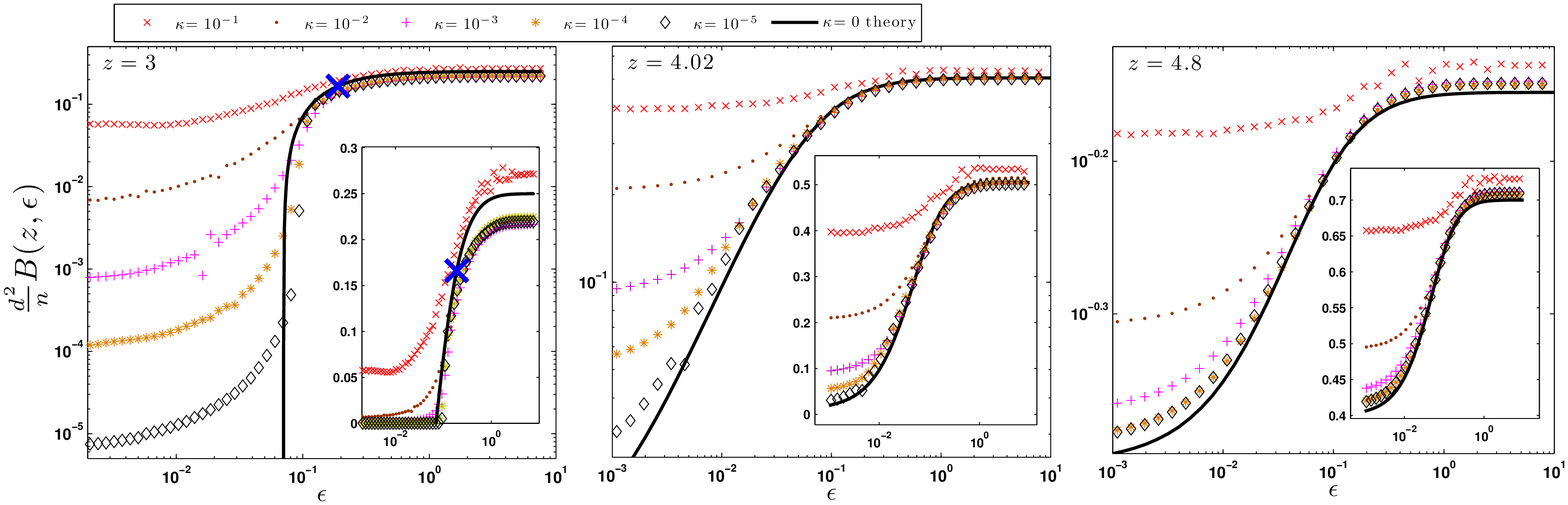}
\par\end{centering}
\caption{The nonlinear bulk modulus for the triangular lattice as a function
of the applied strain for different values of the bending modulus (see legend above the plots). The average coordination numbers are $3$, $4.02$ and $4.8$ (see upper left corner on each plot). The theoretical curves are given by Eq. (\ref{eq:KEM_Diluted}) and, in the explicit form, by Eq. (\ref{eq:K_explicit}). Insets show the same plots on the semi-log scale. Big crosses represent the location of the first order transition as predicted by Eq. (\ref{eq:zc_1})}
\label{Fig3}
\end{figure*}

Many biopolymer networks, including collagen and fibrin networks, have a branched structure with a connectivity  close to three on average~\cite{Lindstrom2010biopolymer}. The rigidity of such networks with connectivities below Maxwell's central-force isostatic point can be accounted for by the existence of additional non-central force interactions such as those arising from fiber bending.
To analyze the effects of the finite fiber bending stiffness on network elasticity we generalize the model presented in \cite{Broedersz2011} to the nonlinear regime. The resulting Hamiltonian is composed of two terms representing the stretching and the bending energies:
\begin{equation}
H=\frac{\mu}{2}\underset{\left\langle ij\right\rangle}{\sum}g_{ij}\left(|\mathbf{u}_{i}-\mathbf{u}_{j}|-1\right)^{2}+\kappa\underset{\left\langle ijk\right\rangle}{\sum}g_{ij}g_{jk}\left(1-\cos \theta_{ijk}\right),
\label{HamAddWeakInt}
\end{equation}
where the summation in the bending term extends
over consecutive  bonds along the same fiber and $\Delta\theta_{ijk}$ is the angle between the $ij$ and the $jk$ bonds. Here, $g_{ij} = 1$ for uncut bonds (with $\mu_{ij}=\mu$) and $g_{ij} = 0$ for bonds that have been cut (with $\mu_{ij}=0$). Thus, in this model the network cross-links are freely hinging.
Various EM theories for this model were developed for the linear elasticity~\cite{das2007effective,Broedersz2011}. However, the generalization to the nonlinear regime seems to be technically challenging. The nonlinear EM theory described above is used to calculate the nonlinear differential bulk modulus in the $\kappa=0$ case. To analyze the importance of the additional interactions we compare our purely central-force, $\kappa=0$, analytical formula Eqs. (\ref{eq:KEM_Diluted},\ref{eq:K_explicit}) for the nonlinear bulk modulus to the numerical results for different values of $\kappa$ (see Fig. \ref{Fig3}). For small enough values of $\kappa$ the nonlinear bulk modulus does not depend on $\kappa$ above the central-force isostatic point. By contrast, below the transition strain, $B\left(\epsilon\right)$ approaches a plateau proportional to $\kappa$, which is not captured by the central-force nonlinear EM theory. However, the strain at which the network exhibits strain stiffening appears to be well approximated by the nonlinear EM theory prediction.
\section{Branched networks and Random Bond Model}
\begin{figure}
\begin{centering}
\includegraphics[clip,width=\columnwidth]{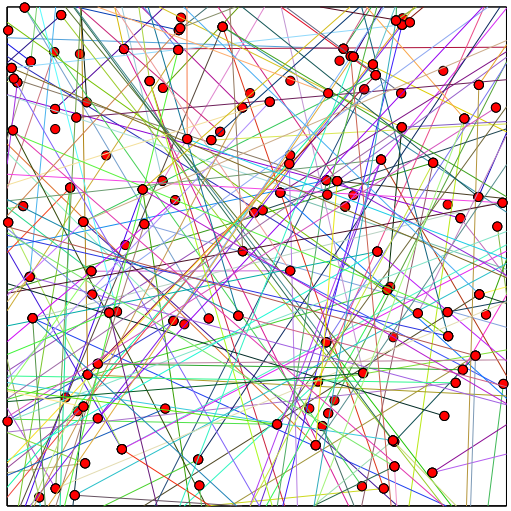}
\par\end{centering}
\caption{Illustration of the Random Bond Model network in two dimensions. Here the density of the bonds is $n_d=1/3$ and the mean coordination number is $z=2$.}
\label{RandomBondIllustration}
\end{figure}

\begin{figure}
\begin{centering}
\includegraphics[clip,width=\columnwidth]{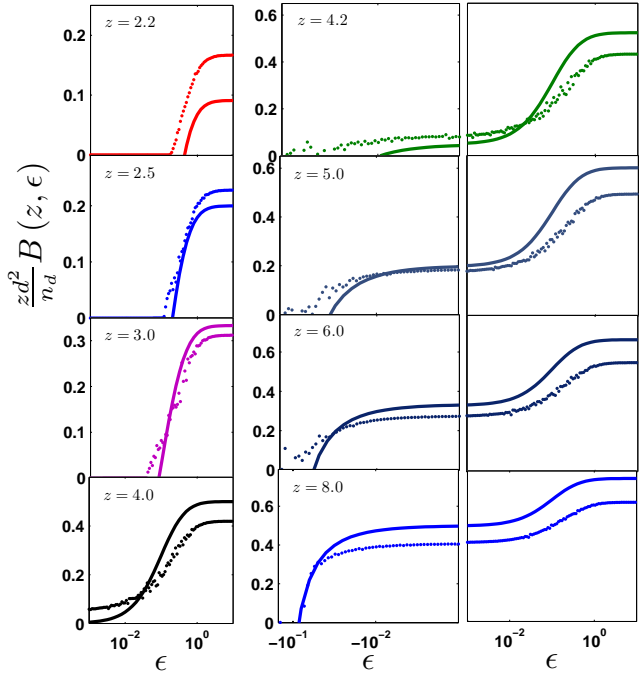}
\par\end{centering}
\caption{The differential nonlinear bulk modulus as a function of the applied strain of the two-dimensional Random Bond Model network for different values of the mean coordination number, $z$. The dots represent the numerical result while the solid lines are based on Eq. (\ref{RandomBondEM}).}
\label{RandomBond}
\end{figure}

The nonlinear EM theory predicts an expression for the bulk modulus (Eq. (\ref{eq:KEM_Diluted}), and more explicitly in Eq. (\ref{eq:K_explicit})) for a regular, diluted elastic network that depends on the geometry of the undiluted network only via its coordination number, $\mathcal{Z}$. This number sets the maximal possible coordination number of the lattice. In some cases, such as collagen-I that exhibits a branched structure \cite{Lindstrom2010biopolymer}, the maximal possible coordination number seems to be very high such that the probability distribution of $z$ is exponential. For such a networks it is instructive to consider a high $\mathcal{Z}$ limit of the expression for the bulk modulus, Eqs. (\ref{eq:KEM_Diluted}) or (\ref{eq:K_explicit}).
More specifically, an effectively off-lattice network with an exponentially distributed coordination number with a given mean
\begin{equation}
z=\mathcal{P}\mathcal{Z}
\end{equation}
can be constructed by almost total dilution, $\mathcal{P} \rightarrow 0$, of a highly coordinated undiluted regular lattice, $\mathcal{Z}\rightarrow\infty$. A particularly simple example of this type of network is the Random Bond Model where randomly located points are randomly connected \cite{jacobs1998comment,thorpe2002generic} (see Fig. \ref{RandomBondIllustration}).
Using this limiting procedure the expressions for the bulk modulus reduces to
\begin{widetext}
\begin{equation}
B_{EM}\left(z,\epsilon\right)=\frac{n_d}{zd^2}\left(z-2-\frac{d-1}{d+2}
\left\{ \frac{2\left(d+1\right)}{\left[1+\left(2+d\right)\epsilon\right]^{3}}-\frac{6}{\left(1+\frac{d+2}{3}\epsilon\right)^{3}}\right\} \right),
\label{RandomBondEM}
\end{equation}
\end{widetext}
where $n_d$ is the density of the bonds of the diluted network.
The transition curve is given by
\begin{equation}
z_c\left(\epsilon\right)=2+\frac{d-1}{d+2}
\left\{ \frac{2\left(d+1\right)}{\left[1+\left(2+d\right)\epsilon\right]^{3}}-\frac{6}{\left(1+\frac{d+2}{3}\epsilon\right)^{3}}\right\}.
\end{equation}
These results may be directly applied to the Random Bond Model. A comparison between the numerical calculation of the Random Bond Model bulk modulus and Eq. (\ref{RandomBondEM}) for $d=2$ is shown in Fig. \ref{RandomBond}.
Branched networks with exponential distribution of the local connectivity are also expected to behave according to Eq. (\ref{RandomBondEM}).

\section{Summary and discussion}
\label{Summary and discussion}
Motivated by the rich nonlinear elastic behavior of disordered materials we studied random spring networks under isotropic strains. We provided a quantitative analytical theory for the strain-stiffening phenomena  that is driven by non-uniform (non-affine) deformation fields originating in the network disorder. We considered disordered networks on lattice topologies of Hookean springs. The disorder is introduced with a non-uniform distribution of the spring constants for the bonds on a regular lattices.
The central parameter that characterizes such networks is the mean coordination number, $z$; the threshold, $z=z_c$, separates a floppy from a rigid phases. However, when some fraction of the bonds are under stress, the isostatic coordination number can shift continuously to lower values \cite{Alexander1998}. This can be realized, for example, by applying a large deformation to the network
\cite{wyart2008} or by introducing local contractile forces \cite{broedersz2011molecular}. It was shown that rigidity can be induced by additional stresses or strains in networks with connectivities below the zero-strain isostatic point, $z_0=z_{c}\left(\epsilon\rightarrow 0\right)=2d$ in $d$ dimensions. As a result, a significant strain stiffening is induced as the network transitions from the floppy to the rigid phase.

Here we have developed a nonlinear effective medium approach for regular central-force networks with
disordered spring constants to provide insight in such behavior. In this model, we expand the Hamiltonian around the affine deformation
state for an isotropically expanded network. Thus, this theory explicitly accounts for non-affine deformations that are small compared to the affinely strained unit cell. The main result of the EM theory approach is the nonlinear differential
bulk modulus given by Eq. (\ref{eq:KEM_Diluted}), where the effective
parameter $\widetilde{\mu}$ may be found for a given spring constant
probability density of the original network, $P\left(\mu_{ij}\right)$,
using Eq. (\ref{eq:mEff_Integral}). We demonstrated that this theory quantitatively captures the nonlinear elastic properties of a bond diluted network for arbitrary strains far from a transition in two and three dimensions.
In particular, this theory predicts a continuous transition curve for the strain dependent isostatic point, varying from $z_0=z_{c}\left(\epsilon\rightarrow 0\right)\simeq2d$ at zero strain to the conductivity threshold $z_{\rm cond}\simeq2$ in the infinite strain limit.  The transition at the strain-dependent rigidity point is accompanied with divergent strain fluctuations, reminiscent of critical behavior. We showed how the nonlinear EM theory can be used to calculate the correlation functions associated to such strain fluctuations. The two point non-affinity parameter, quantifying the relative non-affine deformations of neighboring points, can be used to inspect internal consistency (Ginzburg criterion) of the nonlinear EM theory approach, which breaks down in the vicinity of the (strain dependent) rigidity percolation point.

Application of the EM theory developed here presupposes that the order of the transition is known.
From our numerical results one cannot rule out the possibility of either a first or a second order transition. This remains a subject of further study \cite{sheinman2012NonlinearShear}.

We found that a superisostatic disordered network with $z>2d$ may loose rigidity under positive pressure (negative strain values) in two- and three-dimensional networks. Similar elastic collapse was found and analysed for the perfect triangular lattice \cite{discher1997phase,wintz1997mesh}. Here we showed that the location of such a collapse depends mostly on the network topology via the mean coordination number. The mean-field approach developed here is found to predict reasonably well the location of the collapse and the elastic properties of the network for the negative strain values.

We also investigated the effects of additional weak non-central force interactions numerically in the form of fiber bending in bond diluted networks. The resulting fiber network exhibits a strain stiffening transition from a soft, bending dominated regime to a stretching dominated regime. Importantly, however, this transition still occurs at the transition strain predicted by the central-force nonlinear EM theory, which quantitatively captures the nonlinear elasticity beyond the transition strain. These results may lend insight into the nonlinear elasticity of biological fiber networks.

The EM theory expressions for the elasticity behavior of random spring networks depends on the network geometry only via the coordination number of the undiluted network, $\mathcal{Z}$. One may interpret $\mathcal{Z}$ also as the maximal coordination number of the diluted network. This may lead to the temptation to use the results obtained here for other than diluted regular network systems, including networks with geometrical disorder. However, there is at least one example where a network based on the jammed configuration geometry has qualitatively different elastic behavior than the diluted regular network with the same mean coordination number \cite{ellenbroek2009non}. In the present work we defined the strain of the diluted network relative to the zero energy state of the undiluted network. Therefore, it is unclear how our results may be extrapolated for geometrically disordered networks. Nevertheless, we have shown that our results may be applied to describe the elastic response of the geometrically disordered Random Bond Model.

In this work we focused on the differential \emph{bulk} modulus of networks under strain.
However, for many experimentally relevant systems the \emph{shear} and the \emph{Young's}
moduli may be more relevant. To investigate such systems, a generalization of the nonlinear EM theory presented here
to anisotropic deformations is required. This appears to be technically challenging and will be an interesting subject of further study, along with the order of the transition, transition behavior and various consequences of geometrical and topological disorder.

\begin{acknowledgments}
This work was funded by FOM/NWO and by National Science Foundation under Grant N. NSF PHY05-51164. We thank M. Rubinshtein, M. Das, C. Heusinger and L. Jawerth for fruitful discussions.
We also thank X. Mao and T. C. Lubensky for discussions and for sharing their related work \cite{Mao2010Elasticity}.
\end{acknowledgments}
\appendix
%dummy comment inserted by tex2lyx to ensure that this paragraph is not empty
%dummy comment inserted by tex2lyx to ensure that this paragraph is not empty

\begin{widetext}

\section{The calculation of $\widetilde{\mu}_{EM}$\label{AppA}}

In this Appendix we calculate $\mu_{EM}$---the displacement of the $nm$ bond in the \emph{unperturbed}
EM network due to a unit force $\mathbf{r}_{nm}$ acting on the $nm$
bond.

The dynamical matrix of the unperturbed EM Hamiltonian (\ref{eq:H_EM})
is given by
\begin{equation}
D_{ij}=\begin{cases}
-\frac{\widetilde{\mu}}{1+\epsilon}\left(\mathbf{r}_{ij}\otimes\mathbf{r}_{ij}+\epsilon\mathbb{I}\right) & i\neq j\\
\frac{\widetilde{\mu}}{1+\epsilon}\underset{j\neq i}{\sum}\left(\mathbf{r}_{ij}\otimes\mathbf{r}_{ij}+\epsilon\mathbb{I}\right) & i=j\end{cases},\end{equation}
 where $\mathbb{I}$ is the unit tensor and $\otimes$ is the external
product. The Fourier transform of $D$ is given by
\begin{eqnarray}
D\left(\mathbf{k}\right) & = & \underset{ij}{\sum}D_{ij}e^{i\mathbf{k}\cdot\mathbf{r}_{ij}}=\nonumber \\
 & = & \frac{\widetilde{\mu}}{1+\epsilon}\underset{\mathbf{r}}{\sum}\left(\mathbf{r}\otimes\mathbf{r}+\epsilon\mathbb{I}\right)\left(1-e^{i\mathbf{k}\cdot\mathbf{r}}\right)
\end{eqnarray}
 where $\mathbf{r}$ runs over all unit bond vectors. The unit force
acting on the $nm$ bond is given by
\begin{equation}
\mathbf{f}_{i}=\mathbf{r}_{nm}\left(\delta_{i,n}-\delta_{i,m}\right),
\end{equation}
 so that its Fourier transform is
 \begin{equation}
\mathbf{f}\left(\mathbf{k}\right)=\underset{i}{\sum}\mathbf{f}_{i}e^{i\mathbf{k}\cdot\mathbf{R}_{i}}=\mathbf{r}_{nm}\left(1-e^{i\mathbf{k}\cdot\mathbf{r}_{nm}}\right).
\end{equation}
 Thus the Fourier transform of the displacement field is given by
 \begin{equation}
\mathbf{v}\left(\mathbf{k}\right)=-D^{-1}\left(\mathbf{k}\right)\cdot\mathbf{f}\left(\mathbf{k}\right).
\end{equation}
 The displacement of the $nm$ bond due to the unit force is
%\begin{widetext}
\begin{eqnarray}
\frac{1}{\mu_{EM}} & = & \mathbf{r}_{nm}\cdot\underset{\mathbf{k}}{\sum}\mathbf{v}\left(\mathbf{k}\right)\left(e^{-i\mathbf{k}\cdot\mathbf{r}_{nm}}-1\right)=-\underset{\mathbf{k}}{\sum}\mathbf{r}_{nm}\cdot\mathbf{f}\left(\mathbf{k}\right)D^{-1}\left(\mathbf{k}\right)\left(e^{-i\mathbf{k}\cdot\mathbf{r}_{nm}}-1\right)\nonumber \\
 & = & \frac{1}{\widetilde{\mu}}\frac{2d\left(1+\epsilon\right)}{\mathcal{Z}}\left[1-\frac{\epsilon}{d}\underset{\mathbf{k}}{\sum}Tr\left\{ \frac{\underset{\mathbf{r}}{\sum}\left(1-e^{i\mathbf{k}\cdot\mathbf{r}}\right)}{\underset{\mathbf{r}}{\sum}\left(\mathbf{r}\otimes\mathbf{r}+\epsilon\mathbb{I}\right)\left(1-e^{i\mathbf{k}\cdot\mathbf{r}}\right)}\right\} \right]\equiv\frac{a\left(\epsilon\right)}{\widetilde{\mu}}.
\label{eq:mEM}
\end{eqnarray}
 For a highly coordinated lattice the sum over $\mathbf{r}$ may be
well approximated by the integral over the sphere that includes all
the neighbouring crosslinks and, since the sum over $\mathbf{k}$
is dominated by the small $\mathbf{k}\cdot\mathbf{r\ll1}$
values, $a\left(\epsilon\right)$ may be approximated by
\begin{eqnarray}
a\left(\epsilon\right) & \simeq & \frac{2d\left(1+\epsilon\right)}{\mathcal{Z}}\left[1-\frac{\epsilon}{d}Tr\left\{ \frac{\oint\left(\mathbf{k}\cdot\mathbf{r}\right)^{2}d^{d-1}\mathbf{r}}{\oint\left(\mathbf{k}\cdot\mathbf{r}\right)^{2}d^{d-1}\mathbf{r}\left(\mathbf{r}\otimes\mathbf{r}+\epsilon\mathbb{I}\right)}\right\} \right]=\frac{2d\left(1+\epsilon\right)}{\mathcal{Z}}\left[1-\frac{\epsilon}{d}\left(\frac{1}{\frac{3}{2+d}+\epsilon}+\frac{d-1}{\frac{1}{2+d}+\epsilon}\right)\right].
\label{eq:ad_App}
\end{eqnarray}
%\end{widetext}

\section{Explicit results for diluted networks\label{AppB}}

In this Appendix we present the explicit results for diluted networks.

The nonlinear differential bulk modulus calculated using (\ref{eq:KEM_Diluted})
is given by
%\begin{widetext}
\begin{align}
B_{EM}\left(\epsilon,z\right) & =\frac{n}{d^2}\frac{\mu}{\left((d+2)^{2}(\mathcal{Z}-2)\epsilon^{2}-2\left(d^{2}+7d-2(d+2)\mathcal{Z}+4\right)\epsilon-6d+3\mathcal{Z}\right)^{3}}\times\nonumber \\
\times & \left\{\begin{array}{c}
\epsilon^{3}((z-2)\epsilon((\mathcal{Z}-2)\epsilon((\mathcal{Z}-2)\epsilon-6)+12)-8)d^{6}\\
+2\epsilon^{2}\left(6(z-2)(\mathcal{Z}-2)^{2}\epsilon^{4}+3(z-2)(\mathcal{Z}-2)(2\mathcal{Z}-15)\epsilon^{3}\right)d^{5}\\
+2\epsilon^{2}\left(-3(z-2)(11\mathcal{Z}-42)\epsilon^{2}+4(14z+\mathcal{Z}-39)\epsilon-36\right)d^{5}\\
+\epsilon\left(60(z-2)(\mathcal{Z}-2)^{2}\epsilon^{5}+24(z-2)(\mathcal{Z}-2)(5\mathcal{Z}-21)\epsilon^{4}+3(z-2)(\mathcal{Z}(19\mathcal{Z}-262)+564)\epsilon^{3}\right)d^{4}\\
+\epsilon\left(-2((\mathcal{Z}-334)\mathcal{Z}+z(137\mathcal{Z}-626)+1428)\epsilon^{2}+72(5z+\mathcal{Z}-17)\epsilon-216\right)d^{4}\\
+2\left(80(z-2)(\mathcal{Z}-2)^{2}\epsilon^{6}+24(z-2)(\mathcal{Z}-2)(10\mathcal{Z}-29)\epsilon^{5}\right)d^{3}\\
+2\left(12(z-2)(\mathcal{Z}(19\mathcal{Z}-140)+206)\epsilon^{4}+((2398-167\mathcal{Z})\mathcal{Z}\right)d^{3}\\
+2\left(z(\mathcal{Z}(68\mathcal{Z}-1063)+2198)-4636)\epsilon^{3}-9((\mathcal{Z}-94)\mathcal{Z}\right)d^{3}\\
+2\left(z(29\mathcal{Z}-114)+276)\epsilon^{2}+108(2z+\mathcal{Z}-7)\epsilon-108\right)d^{3}\\
+\left(240(z-2)(\mathcal{Z}-2)^{2}\epsilon^{6}+96(z-2)(\mathcal{Z}-2)(10\mathcal{Z}-21)\epsilon^{5}+24(z-2)(\mathcal{Z}(57\mathcal{Z}-274)+282)\epsilon^{4}\right)d^{2}\\
+\left(9(12(51-5\mathcal{Z})\mathcal{Z}+z(19(\mathcal{Z}-14)\mathcal{Z}+336)-544)\epsilon^{2}-54(8z(\mathcal{Z}-2)+(\mathcal{Z}-28)\mathcal{Z}+16)\epsilon+108(z+2\mathcal{Z})\right)d^{2}\\
+\left(8(5(286-45\mathcal{Z})\mathcal{Z}+z(3\mathcal{Z}(34\mathcal{Z}-231)+742)-1428)\epsilon^{3}\right)d^{2}\\
+2\left(96(z-2)(\mathcal{Z}-2)^{2}\epsilon^{6}+240(z-2)(\mathcal{Z}(2\mathcal{Z}-7)+6)\epsilon^{5}\right)d\\
+2\left(48(z-2)(\mathcal{Z}-1)(19\mathcal{Z}-42)\epsilon^{4}+4((1262-407\mathcal{Z})\mathcal{Z}\right)d\\
+2\left(z(\mathcal{Z}(204\mathcal{Z}-679)+334)-624)\epsilon^{3}+18(z(\mathcal{Z}(19\mathcal{Z}-80)+20)\right)d\\
+2\left(-2(3\mathcal{Z}-8)(7\mathcal{Z}-2))\epsilon^{2}+27(2z(\mathcal{Z}-8)-7\mathcal{Z}+16)\mathcal{Z}\epsilon-27\mathcal{Z}(2z+\mathcal{Z})\right)d\\
+z\left(64(4\epsilon(\epsilon(\epsilon+3)+3)+5)\epsilon^{3}-8\mathcal{Z}(4\epsilon(\epsilon(4\epsilon(2\epsilon+9)+57)+41)+45)\epsilon^{2}+\mathcal{Z}^{2}(4\epsilon(\epsilon+2)+3)^{3}\right)\\
-8\epsilon\left(64\epsilon^{2}(\epsilon+1)^{3}-8\mathcal{Z}\epsilon(\epsilon(\epsilon(4\epsilon(2\epsilon+9)+57)+37)+9)+\mathcal{Z}^{2}(2\epsilon(2\epsilon(\epsilon(4\epsilon(\epsilon+6)+57)+61)+63)+27)\right)\end{array}\right\} \label{eq:K_explicit}
\end{align}
for $\epsilon>\epsilon_c$. Below the transition curve the nonlinear differential bulk modulus vanishes.
The transition curve for the assumption of the first order transition is given by
\begin{equation}
z_{c_{1}}\left(\epsilon\right)=\frac{\left\{ \begin{array}{c}
18d(-2d+\mathcal{Z})+3(-8d(4+d(7+d))+(12+d(29+7d))\mathcal{Z})\epsilon\\
-4(16+d(80+d(81+d(20+d-2\mathcal{Z})-21\mathcal{Z})-48\mathcal{Z})-28\mathcal{Z})\epsilon^{2}\\
+(2+d)^{2}(-8(4+d(7+d))+(28+d(19+d))\mathcal{Z})\epsilon^{3}+2(2+d)^{4}(-2+\mathcal{Z})\epsilon^{4}\end{array}\right\} }{\left\{ \begin{array}{c}
9(-2d+\mathcal{Z})-3\left(4+31d+13d^{2}-8(2+d)\mathcal{Z}\right)\epsilon-2(2+d)(2(8+5d(4+d))-11(2+d)\mathcal{Z})\epsilon^{2}\\
-(2+d)^{2}\left(20+25d+3d^{2}-8(2+d)\mathcal{Z}\right)\epsilon^{3}+(2+d)^{4}(-2+\mathcal{Z})\epsilon^{4}\end{array}\right\} }.
\label{eq:zc_1}
\end{equation}
The transition curve for the assumption of the second order transition is given by
\begin{equation}
z_{c_2}\left(\epsilon\right)=2\frac{\left\{ \begin{array}{c}
(d+2)^{6}(\mathcal{Z}-2)^{2}\epsilon^{6}-6(d+2)^{4}(\mathcal{Z}-2)\left(d^{2}+7d-2(d+2)\mathcal{Z}+4\right)\epsilon^{5}\\
+3(d+2)^{2}\left(19(d+2)^{2}\mathcal{Z}^{2}-2(d+2)(d(11d+65)+38)\mathcal{Z}+4(d(d+4)(d(d+13)+17)+16)\right)\epsilon^{4}\\
+\left((d+2)^{2}(d(d+163)+244)\mathcal{Z}^{2}-2(d+2)(d(d(d(2d+163)+873)+1114)+296)\mathcal{Z}\right)\epsilon^{3}\\
+\left(4(d(d+7)+4)(d(d(d(d+32)+129)+128)+16)\right)\epsilon^{3}\\
+9(2d-\mathcal{Z})(2(d(d+4)(d(d+13)+17)+16)-(d+2)(d(d+28)+28)\mathcal{Z})\epsilon^{2}\\
+27(d(d+7)+4)(\mathcal{Z}-2d)^{2}\epsilon+27d(\mathcal{Z}-2d)^{2}\end{array}\right\} }{\left\{ \begin{array}{c}
(d+2)^{6}(\mathcal{Z}-2)^{2}\epsilon^{6}-6(d+2)^{4}(\mathcal{Z}-2)\left(d^{2}+7d-2(d+2)\mathcal{Z}+4\right)\epsilon^{5}\\
+3(d+2)^{2}\left(19(d+2)^{2}\mathcal{Z}^{2}-2(d+2)(d(11d+65)+38)\mathcal{Z}+4(d(d+4)(d(d+13)+17)+16)\right)\epsilon^{4}\\
+2(d+2)\left(68(d+2)^{2}\mathcal{Z}^{2}-(d+2)(d(137d+515)+164)\mathcal{Z}+2d(d(d+5)(28d+117)+314)+80\right)\epsilon^{3}\\
+9(d+2)(d(20d-19\mathcal{Z}+74)-38\mathcal{Z}+20)(2d-\mathcal{Z})\epsilon^{2}+108(d+2)(\mathcal{Z}-2d)^{2}\epsilon+27(\mathcal{Z}-2d)^{2}\end{array}\right\} }.
\label{eq:zc_2}
\end{equation}
%\end{widetext}

\section{Calculation of the differential non-affinity parameter}
\label{CalcDiffNonAffinity}
The only unknown term in the differential non-affinity expression in Eq. (\ref{DiffNonAff}) is $\left\langle \left(\frac{d\mathbf{v}_{n}\left(\epsilon\right)}{d\epsilon}\right)^{2}\right\rangle$. This term may be evaluated within the framework of the EM theory as following. Using Eqs. (\ref{ForceVanishes},\ref{ForcesUncorrelated},\ref{DispToForce}) one obtains
%\begin{widetext}
\begin{equation}
\left\langle \left(\frac{d\mathbf{v}_{n}\left(\epsilon\right)}{d\epsilon}\right)^{2}\right\rangle =\frac{1}{2N^{2}}\underset{\mathbf{r},\mathbf{k}}{\sum}\left\{ -\begin{array}{c}
\left\langle \left\{ \frac{d}{d\epsilon}\left[\frac{\epsilon\left(1+\epsilon\right)}{a\left(\epsilon\right)}\frac{\widetilde{\mu}-\mu_{ij}}{\frac{\widetilde{\mu}}{a\left(\epsilon\right)}-\widetilde{\mu}+\mu_{ij}}\right]\right\} ^{2}\right\rangle \frac{\left(1-e^{i\mathbf{k}\cdot\mathbf{r}}\right)}{\underset{\mathbf{r}}{\sum}\left(\mathbf{r}\otimes\mathbf{r}+\epsilon\mathbb{I}\right)\left(1-e^{i\mathbf{k}\cdot\mathbf{r}}\right)}\mathbf{r}\frac{\left(1-e^{-i\mathbf{k}\cdot\mathbf{r}}\right)}{\underset{\mathbf{r}}{\sum}\left(\mathbf{r}\otimes\mathbf{r}+\epsilon\mathbb{I}\right)\left(1-e^{-i\mathbf{k}\cdot\mathbf{r}}\right)}\mathbf{r}\\
\left\langle \frac{d}{d\epsilon}\left[\frac{\epsilon\left(1+\epsilon\right)}{a\left(\epsilon\right)}\frac{\widetilde{\mu}-\mu_{ij}}{\frac{\widetilde{\mu}}{a\left(\epsilon\right)}-\widetilde{\mu}+\mu_{ij}}\right]^{2}\right\rangle \frac{\left(1-e^{i\mathbf{k}\cdot\mathbf{r}}\right)}{\underset{\mathbf{r}}{\sum}\left(\mathbf{r}\otimes\mathbf{r}+\epsilon\mathbb{I}\right)\left(1-e^{i\mathbf{k}\cdot\mathbf{r}}\right)}\mathbf{r}\frac{\left(1-e^{-i\mathbf{k}\cdot\mathbf{r}}\right)\underset{\mathbf{r'}}{\sum}\left(1-e^{-i\mathbf{k}\cdot\mathbf{r}'}\right)}{\left[\underset{\mathbf{r}}{\sum}\left(\mathbf{r}\otimes\mathbf{r}+\epsilon\mathbb{I}\right)\left(1-e^{-i\mathbf{k}\cdot\mathbf{r}}\right)\right]^{2}}\mathbf{r}\\
\left\langle \left[\frac{\epsilon\left(1+\epsilon\right)}{a\left(\epsilon\right)}\frac{\widetilde{\mu}-\mu_{ij}}{\frac{\widetilde{\mu}}{a\left(\epsilon\right)}-\widetilde{\mu}+\mu_{ij}}\right]^{2}\right\rangle \frac{\left(1-e^{i\mathbf{k}\cdot\mathbf{r}}\right)\underset{\mathbf{r'}}{\sum}\left(1-e^{-i\mathbf{k}\cdot\mathbf{r}'}\right)}{\left[\underset{\mathbf{r}}{\sum}\left(\mathbf{r}\otimes\mathbf{r}+\epsilon\mathbb{I}\right)\left(1-e^{i\mathbf{k}\cdot\mathbf{r}}\right)\right]^{2}}\mathbf{r}\frac{\left(1-e^{-i\mathbf{k}\cdot\mathbf{r}}\right)\underset{\mathbf{r'}}{\sum}\left(1-e^{-i\mathbf{k}\cdot\mathbf{r}'}\right)}{\left[\underset{\mathbf{r}}{\sum}\left(\mathbf{r}\otimes\mathbf{r}+\epsilon\mathbb{I}\right)\left(1-e^{-i\mathbf{k}\cdot\mathbf{r}}\right)\right]^{2}}\mathbf{r}\end{array}\right\} .
\end{equation}
%\end{widetext}
As before we apply the approximation of the highly
coordinated lattice and get
%\begin{widetext}
\begin{equation}
\left\langle \left(\frac{d\mathbf{v}_{n}\left(\epsilon\right)}{d\epsilon}\right)^{2}\right\rangle =\frac{d}{2\mathcal{Z}}A_{d}f_{d}\left(N\right)\left\{ -\begin{array}{c}
\left\langle \left\{ \frac{d}{d\epsilon}\left[\frac{\epsilon\left(1+\epsilon\right)}{a\left(\epsilon\right)}\frac{\widetilde{\mu}-\mu_{ij}}{\frac{\widetilde{\mu}}{a\left(\epsilon\right)}-\widetilde{\mu}+\mu_{ij}}\right]\right\} ^{2}\right\rangle \left\{ \frac{\frac{3}{2+d}}{\left(\frac{3}{2+d}+\epsilon\right)^{2}}+\frac{\frac{d-1}{2+d}}{\left(\frac{1}{2+d}+\epsilon\right)^{2}}\right\} -\\
\frac{d}{d\epsilon}\left\langle \left[\frac{\epsilon\left(1+\epsilon\right)}{a\left(\epsilon\right)}\frac{\widetilde{\mu}-\mu_{ij}}{\frac{\widetilde{\mu}}{a\left(\epsilon\right)}-\widetilde{\mu}+\mu_{ij}}\right]^{2}\right\rangle \left\{ \frac{\frac{3}{2+d}}{\left(\frac{3}{2+d}+\epsilon\right)^{3}}+\frac{\frac{d-1}{2+d}}{\left(\frac{1}{2+d}+\epsilon\right)^{3}}\right\} +\\
+\left\langle \left[\frac{\epsilon\left(1+\epsilon\right)}{a\left(\epsilon\right)}\frac{\widetilde{\mu}-\mu_{ij}}{\frac{\widetilde{\mu}}{a\left(\epsilon\right)}-\widetilde{\mu}+\mu_{ij}}\right]^{2}\right\rangle \left\{ \frac{\frac{3}{2+d}}{\left(\frac{3}{2+d}+\epsilon\right)^{4}}+\frac{\frac{d-1}{2+d}}{\left(\frac{1}{2+d}+\epsilon\right)^{4}}\right\} \end{array}\right\}
\end{equation}
%\end{widetext}
\end{widetext}
\bibliographystyle{apsrev}
\bibliography{NonLinear}
\end{document}